\documentclass[iop]{emulateapj}
\usepackage{graphicx}
\usepackage{enumerate}
\usepackage{amssymb, amsmath}
\usepackage{natbib}
\usepackage{color}
\usepackage{ulem}
\usepackage{dblfnote}
\usepackage{appendix}
\usepackage{hyperref}
\usepackage[stable]{footmisc}

\definecolor{red}{rgb}{1.0,0.0,0.0}

\def\pasa{PASA}
\def\procspie{Proc.~SPIE}

\newcommand{\ipac}{1}
\newcommand{\nexsci}{2}
\newcommand{\tokyo}{3}
\newcommand{\naoj}{4}
\newcommand{\ames}{5}
\newcommand{\subaru}{6}
\newcommand{\eureka}{7}
\newcommand{\abc}{8}
\newcommand{\stanford}{9}
\newcommand{\ucsb}{10}
\newcommand{\arizona}{11}
\newcommand{\caltech}{12}
\newcommand{\cotedazur}{13}
\newcommand{\princetoneng}{14}
\newcommand{\goddard}{15}
\newcommand{\notredame}{16}
\newcommand{\stockholm}{17}
\newcommand{\princetonastro}{18}
\newcommand{\jpl}{19}

\begin{document}

\author{Taichi Uyama$^{\ipac,\nexsci,\tokyo,\naoj}$}
\author{Thayne Currie$^{\ames,\subaru,\eureka}$}
\author{Yasunori Hori$^{\abc,\naoj}$}
\author{Robert J. De Rosa$^{\stanford}$}
\author{Kyle Mede$^{\tokyo}$}
\author{Timothy D. Brandt$^{\ucsb}$}
\author{Jungmi Kwon$^{\tokyo}$}
\author{Olivier Guyon$^{\subaru,\arizona,\abc}$}
\author{Julien Lozi$^{\subaru}$}
\author{Nemanja Jovanovic$^{\caltech}$}
\author{Frantz Martinache$^{\cotedazur}$}
\author{Tomoyuki Kudo$^{\subaru}$}
\author{Motohide Tamura$^{\tokyo,\abc,\naoj}$}
\author{N. Jeremy Kasdin$^{\princetoneng}$}
\author{Tyler Groff$^{\goddard}$}
\author{Jeffrey Chilcote$^{\notredame}$}
\author{Masahiko Hayashi$^{\naoj}$}
\author{Michael W. McElwain$^{\goddard}$}
\author{Ruben Asensio-Torres$^{\stockholm}$}
\author{Markus Janson$^{\stockholm}$}
\author{Gillian R. Knapp$^{\princetonastro}$}
\author{Eugene Serabyn$^{\jpl}$}

\footnotetext[\ipac]{Infrared Processing and Analysis Center, California Institute of Technology, Pasadena, CA 91125, USA}
\footnotetext[\nexsci]{NASA Exoplanet Science Institute}
\footnotetext[\tokyo]{Department of Astronomy, The University of Tokyo, 7-3-1, Hongo, Bunkyo-ku, Tokyo 113-0033, Japan}
\footnotetext[\naoj]{National Astronomical Observatory of Japan, 2-21-1 Osawa, Mitaka, Tokyo 181-8588, Japan}
\footnotetext[\ames]{NASA-Ames Research Center, Moffett Field, CA, USA}
\footnotetext[\subaru]{Subaru Telescope, National Astronomical Observatory of Japan, 650 North A‘ohoku Place, Hilo, HI96720, USA}
\footnotetext[\eureka]{Eureka Scientific, 2452 Delmer Street Suite 100, Oakland, CA, USA}
\footnotetext[\abc]{Astrobiology Center of NINS, 2-21-1 Osawa, Mitaka, Tokyo 181-8588, Japan}
\footnotetext[\stanford]{Kavli Institute for Particle Astrophysics and Cosmology, Stanford University, Stanford, CA 94305, USA}
\footnotetext[\ucsb]{Department of Physics, University of California, Santa Barbara, Santa Barbara, California, USA}
\footnotetext[\arizona]{Steward Observatory, University of Arizona, Tucson, AZ 85721, USA}
\footnotetext[\caltech]{Department of Astronomy, California Institute of Technology, 1200 E. California Blvd., Pasadena, CA 91125, USA}
\footnotetext[\cotedazur]{Universit$\acute{e}$ C$\hat{o}$te d'Azur, Observatoire de la C$\hat{o}$te d'Azur, CNRS, Laboratoire Lagrange, France}
\footnotetext[\princetoneng]{Department of Mechanical Engineering, Princeton University, Princeton, NJ, USA}
\footnotetext[\goddard]{NASA-Goddard Space Flight Center, Greenbelt, MD, USA}
\footnotetext[\notredame]{Department of Physics, University of Notre Dame, South Bend, IN, USA}
\footnotetext[\stockholm]{Department of Astronomy, Stockholm University, AlbaNova University Center, SE-106 91 Stockholm, Sweden}
\footnotetext[\princetonastro]{Department of Astrophysical Science, Princeton University, Peyton Hall, Ivy Lane, Princeton, NJ 08544, USA}
\footnotetext[\jpl]{Jet Propulsion Laboratory, California Institute of Technology, Pasadena, CA, 91109, USA}

\shortauthors{Uyama et al.}

\title{Atmospheric Characterization and Further Orbital Modeling of $\kappa$ And b}

\begin{abstract}
We present $\kappa$ Andromeda b's photometry and astrometry taken with Subaru/SCExAO+HiCIAO and Keck/NIRC2, combined with recently published SCExAO/CHARIS low-resolution spectroscopy and published thermal infrared photometry to further constrain the companion's atmospheric properties and orbit.  $\kappa$ And b's Y/Y-K colors are redder than field dwarfs, consistent with its youth and lower gravity.
Empirical comparisons of its Y-band photometry and CHARIS spectrum to a large spectral library of isolated field dwarfs reaffirm the conclusion from Currie et al. (2018) that it likely has a low gravity but admit a wider range of most plausible spectral types (L0-L2). Our gravitational classification also suggests that the best-fit objects for $\kappa$ And b may have lower gravity than those previously reported.
Atmospheric models lacking dust/clouds fail to reproduce its entire 1--4.7 $\mu m$ spectral energy distribution, cloudy atmosphere models with temperatures of $\sim$ 1700--2000 $K$ better match $\kappa$ And b data.  Most well-fitting model comparisons favor 1700--1900 $K$, a surface gravity of log(g) $\sim$ 4--4.5, and a radius of 1.3--1.6\,$R_{\rm Jup}$; the best-fit model (DRIFT-Phoenix) yields the coolest and lowest-gravity values: $T_{\rm eff}$=1700 K and $\log g$=4.0.
An update to $\kappa$ And b's orbit with ExoSOFT using new astrometry spanning seven years reaffirms its high eccentricity ($0.77\pm0.08$).
We consider a scenario where unseen companions are
responsible for scattering $\kappa$ And b to a wide separation and high eccentricity.  If three planets, including $\kappa$ And b, were born with coplanar orbits and one of them was ejected by gravitational scattering, a potential inner companion with mass $\gtrsim10M_{\rm Jup}$ could be located at $\lesssim$ 25 au.

\end{abstract}

\section{Introduction}
With the development of Adaptive Optics (AO), direct imaging has uniquely been probing exoplanet populations of young and wide-orbit gas giants \citep[e.g., ][]{Marois2010,Lagrange2010,Kuzuhara2013,Currie2015,Macintosh2015,Keppler2018}.
Young gas giants are amenable to direct detection at infrared wavelength as they are still radiating away their heat of formation, which means that these planets still have vestiges of planet formation, and are bright enough to be resolved with high-contrast imaging instruments around nearby, bright stars.

Photometric and low-resolution spectroscopic measurements of directly-imaged planets and young substellar objects can be used to estimate bulk atmospheric properties.   Broadband (spectro-)photometry over a wide wavelength range can reveal young planet/brown dwarf atmospheres that are cloudier and/or dustier than isolated field substellar objects of the same temperatures \citep{Currie2011,Currie2013,Liu2013,DeRosa2016,Rajan2017}.   Spectral shapes in the major near-IR passbands can diagnose evidence for low surface gravity in young objects \citep{Kirkpatrick2006,Allers2013,Currie2014b}.   Well-calibrated, high signal-to-noise ratio spectra for isolated young and field brown dwarfs can help constrain the spectral type and gravity classification of directly-imaged exoplanets \citep[e.g.][]{Bonnefoy2016,Chilcote2017,Currie2018}.   Atmospheric modeling provides a constraint on the temperature, cloud structure, luminosity and (possibly) gravity of imaged exoplanets \citep[e.g.][]{Currie2011,Barman2015,Chilcote2017,Rajan2017}.

Previous studies have shown that {\it in-situ} core accretion \citep{Pollack1996} or gravitational instability \citep{Boss2011} scenarios struggle to reproduce mass-semimajor axis distributions of the observed planets beyond $\sim 10$\,au \citep[e.g.,][]{Boley2009, Currie2011}.
Therefore, gravitational scattering between planets is proposed to assist formation of wide-orbit planets in the core accretion process \citep[e.g.,][]{2002Icar..156..570M,2008ApJ...686..621F,Nagasawa2008}.
Detecting counterparts that were involved in planet-planet scattering, however, are elusive.
The number of confirmed directly-imaged planets ($\sim$ 10-20) impedes our ability to constrain their formation and evolution scenarios; current frequencies of giant planets beyond $\sim 10$\,au derived from direct imaging surveys are $\sim$10\% or less \citep[e.g.,][]{Brandt2014,Bowler2016,Uyama2017,Nielsen2019}.
Thus, continuous efforts to directly image and characterize wide-orbit planetary systems around young stars are essential to understand the formation mechanisms of wide-orbit planets.


In this study, we target a bright, young, and nearby B9V star, $\kappa$ And (see Table \ref{Kappa And parameters} for the stellar parameters). The Strategic Explorations of Exoplanets and Disks with Subaru \citep[SEEDS;][]{Tamura2009} reported that $\kappa$ And harbors a substellar-mass companion \citep[$\kappa$ And b; ][]{Carson2013}. While early studies admit a wide range of potential ages for the system \citep{Carson2013,Bonnefoy:2014dx, Hinkley2013}, follow-up studies showed that the system is young, with a likely age of $\sim$ 40 $Myr$ \citep{Jones2016} and kinematics that might be consistent with membership in the $\sim$ 20--50 Myr old Columba association \citep[][]{Currie2018}.    Early spectral energy distribution modeling of $\kappa$ And b's photometry from $J$ through $M\prime$ (1.25--5 $\mu m$) suggested the companion had a temperature of 1700--2000 $K$ but could not constrain its surface gravity \citep{{Bonnefoy:2014dx}}.  Near-infrared Subaru/SCExAO+CHARIS spectroscopy of $\kappa$ And b from \citet{Currie2018} showed that the companion was well-matched to low gravity, L0--L1 spectral templates and free-floating substellar objects, with an implied mass of 13$^{+12}_{-2}$ $M_{\rm Jup}$.   Spectral energy distribution modeling of $\kappa$ And b over a wide wavelength range and incorporating both near-IR spectroscopy and photometry, allows us to revisit estimates of its temperature, better constrain its atmospheric properties (e.g. clouds), and potentially quantify its surface gravity.

Characterizing $\kappa$ And b may provide broader insights into the nature of a new class of directly imaged companions.   The estimated semimajor axis of $\kappa$ And b ($\sim$ 55--125 au)  places it at a separation where formation by core accretion is difficult, yet its orbital inclination may imply formation in a disk, perhaps by disk instability \citep[][]{Currie2018}.  Other recent high-contrast imaging studies have also reported substellar-mass companions at these separations around B and early A-type stars with masses nominally above the deuterium-burning limit \citep[e.g., HIP 64892 and HIP 79098;][]{Cheetman2018, Janson2019} and below it \citep[HIP 65426;][]{Chauvin2017}.   In addition to atmospheric characterization, improved orbital measurements of $\kappa$ And b could better constrain its eccentricity, semimajor axis, and alignment with the star's rotation axis.  

Here we aim at updating characterizations of the $\kappa$ And system by using Subaru/HiCIAO+SCExAO and Keck/NIRC2 (Section \ref{sec: Observations and Results 2}).
By expanding wavelength coverage for $\kappa$ And b, we perform a more robust comparison with other substellar objects and synthetic atmospheric models, allowing us to better constrain the companion's temperature and gravity and infer its cloud properties (Sections 3 and 4).   Additionally, we expand the planet's astrometric coverage, adding two additional epochs to update an estimate of its orbital properties (Section 5).   We discuss possible formation and evolution scenarios accounting for $\kappa$ And b's properties in Section 6.

\begin{table}
  \caption{Adopted stellar parameters for $\kappa$ And}
  \centering
  \begin{tabular}{ccc}\\ \hline\hline
  parameters & $\kappa$ And &  Ref. \\ \hline
  RA & 23:40:24.506 & a \\ 
  Dec & +44:20:02.18 & a \\
  Sp type & B9 &  b,c \\
  Mass [$M_{\odot}$] & 2.6--2.8 & b,c,d \\
  Age [Myr] & $47^{+27}_{-40}$ & b \\
  Distance [pc] & 50.0$\pm$0.1 & a \\ \hline
  \end{tabular}
  \footnotetext[1]{\cite{GaiaDR2-2018}}
  \footnotetext[2]{\cite{Jones2016}} 
  \footnotetext[3]{\cite{Currie2018}} 
  \footnotetext[4]{\cite{Bonnefoy:2014dx}}
  \label{Kappa And parameters}
\end{table}

\section{Data} \label{sec: Observations and Results 2}
\subsection{Observations}
\subsubsection{Subaru/SCExAO+HiCIAO}
$\kappa$ Andromedae was observed on UT 18 July 2016 with SCExAO coupled to the HiCIAO infrared camera operating in the Mauna Kea H ($\lambda$ = 1.49--1.78 $\mu$m) and Y (0.957--1.120 $\mu$m) broadband filters (Table \ref{logs}) with a pixel scale of 0\farcs{}0083 pixel$^{-1}$.
Conditions were photometric and slightly above-average in quality for Maunakea: visual seeing of 0\farcs{}4--0\farcs{}5, negligible humidity, and light winds (2 m s$^{-1}$).   


In both filters, science frames consisted of 30-second coadded exposures (Six coadds of 5-second individual frames).  As we did not use a coronagraph in either case, the primary star halo is saturated out to $\rho$ $\sim$ 0\farcs{}2--0\farcs{}25 and 0\farcs{}15--0\farcs{}2 in $H$- and $Y$-band, respectively. We also took unsaturated images in both bands for point spread function (PSF) reference with 5-second integration time and ND0.1 filter. 
Measured full width at half maximum (FWHM) in both sets of unsaturated frames is 5.2 pixels in the $H$-band and 6.2 pixels in the $Y$-band, respectively. 

We utilized angular differential imaging \citep[ADI;][]{Marois2006} to achieve high contrast enough to detect fainter objects around the central star, yielding significantly.   Our field rotation due to ADI ($\sim$ 41--42$^{o}$) is larger and integration time ($t_{int}$ $\sim$ 25-30 minutes) is greater than the higher quality SCExAO/CHARIS data presented in \citet{Currie2018} (10.5$^{o}$ rotation, 14.4 minutes of integration time).
  
The transmission of each ND0.1 filter was measured after the observations to be 0.0085$\pm$0.0006\% in $Y$-band and 0.063$\pm$0.020\% in $H$-band.
We found that the $H$-band ND0.1 filter has large uncertainty and thus we cannot conduct accurate relative photometry using $\kappa$ And A.
Therefore we alternatively used unsaturated images of HIP 79977, which were taken in the same epoch with the $H$-band ND1 filter (0.854$\pm$0.002\%), as photometric reference. 
For Y band, because $\kappa$ And A lacks published precise Y band photometry, we also took unsaturated frames of HIP 118133, as a photometric reference, with 5-second integration time and the $Y$-band ND1 filter (0.388$\pm$0.008\%).
Detailed discussions of photometry are given in Section \ref{sec: Photometry and Astrometry}.

We also took advantages of a SCExAO engineering data set taken in 2 August 2015.
The inaccurate ND0.1 filter was also mainly used for unsaturated frames of $\kappa$ And in this epoch and we used one unsaturated frame, with which the $H$-band ND1 filter was used, for a photometric reference. 
Furthermore, this epoch did not take a globular cluster or a binary system for distortion correction, which yields a systematic astrometry offset.
Although we report our results of photometry and astrometry, we do not use the astrometric result of this engineering run for the discussion hereafter.

\subsubsection{Keck/NIRC2}
To add new constraints on $\kappa$ And b's orbit, we obtained follow-up observations of $\kappa$ And with Keck/NIRC2 in the $K_{\rm s}$-band ($\lambda=1.99-2.30 \mu{\rm m}$ filter using the Lyot coronagraph with a 400 mas occulting spot.

\begin{table*}
\caption[Observing logs for $\kappa$ And]{Observing logs for $\kappa$ And}
\centering
\begin{tabular}{cccccc} \\ \hline\hline
Date ($HST$)& instrument & Band & $T_{\rm exp}$ [min] & Rotation Angle [deg] & remarks \\ \hline
   2015-08-02 & Subaru/HiCIAO+SCExAO & $H$ & 35.0 & 27.70 & SCExAO engineering obs \\
   2016-07-18 & Subaru/HiCIAO+SCExAO & $H$ & 25.0 & 41.70 & science obs  \\ 
   2016-07-18 & Subaru/HiCIAO+SCExAO & $Y$ & 30.5 & 41.31 & science obs for photometry  \\
   2018-11-01 & Keck/NIRC2 & $K_{\rm s}$ & 10 &3.70& science obs for astrometry \\ \hline
\end{tabular}
\label{logs}
\end{table*}

\subsection{Data Reduction}
Basic imaging processing -- e.g. flat fielding, dark subtraction, badpixel mask, distortion correction, and precise PSF registration -- followed previous methods taken for SCExAO/HiCIAO data \citep[][]{Garcia2017,Currie2017}.   
In the distortion correction we used a master distortion map of SCExAO+HiCIAO, which is made by observing a globular cluster of M15 \citep[][]{Currie2017}. 
Registered images were visually inspected to identify a few with poorer AO correction and/or data transfer errors from HiCIAO (e.g., sporadic NaN stripes in one or two channels).

For point-spread function (PSF) subtraction of the HiCIAO data sets, we used a slightly modified version of the \textit{locally optimized combination of images} (LOCI) pipeline \citep[LOCI;][]{Lafreniere2007}, inverting the covariance matrix in LOCI using truncated singular value decomposition (SVD) as in A-LOCI \citep{Currie2012c,Currie2019}.  As $\kappa$ And b is visible in the raw $H$-band data, we opted for conservative settings for both filters: a rotation gap of 0.75 $\lambda$/D, an optimization zone from which we constructed a weighted reference PSF of 300 PSF footprints, and a light SVD cutoff of 10$^{-7}$.

For the Keck/NIRC2 coronagraphic data, basic image processing followed previous methods \citep[e.g.][]{Currie2012b}.   Briefly, after applying corrections for linearity, dark subtraction, and flat-fielding, we registered the images to a common center using stellar PSF seen through the partially transmissive mask.   For PSF subtraction, we used A-LOCI with local masking and a singular value decomposition cutoff of 10$^{-6}$.

\begin{figure*}
\begin{tabular}{cc}
\begin{minipage}{0.5\hsize}
\centering
\includegraphics[scale=0.35]{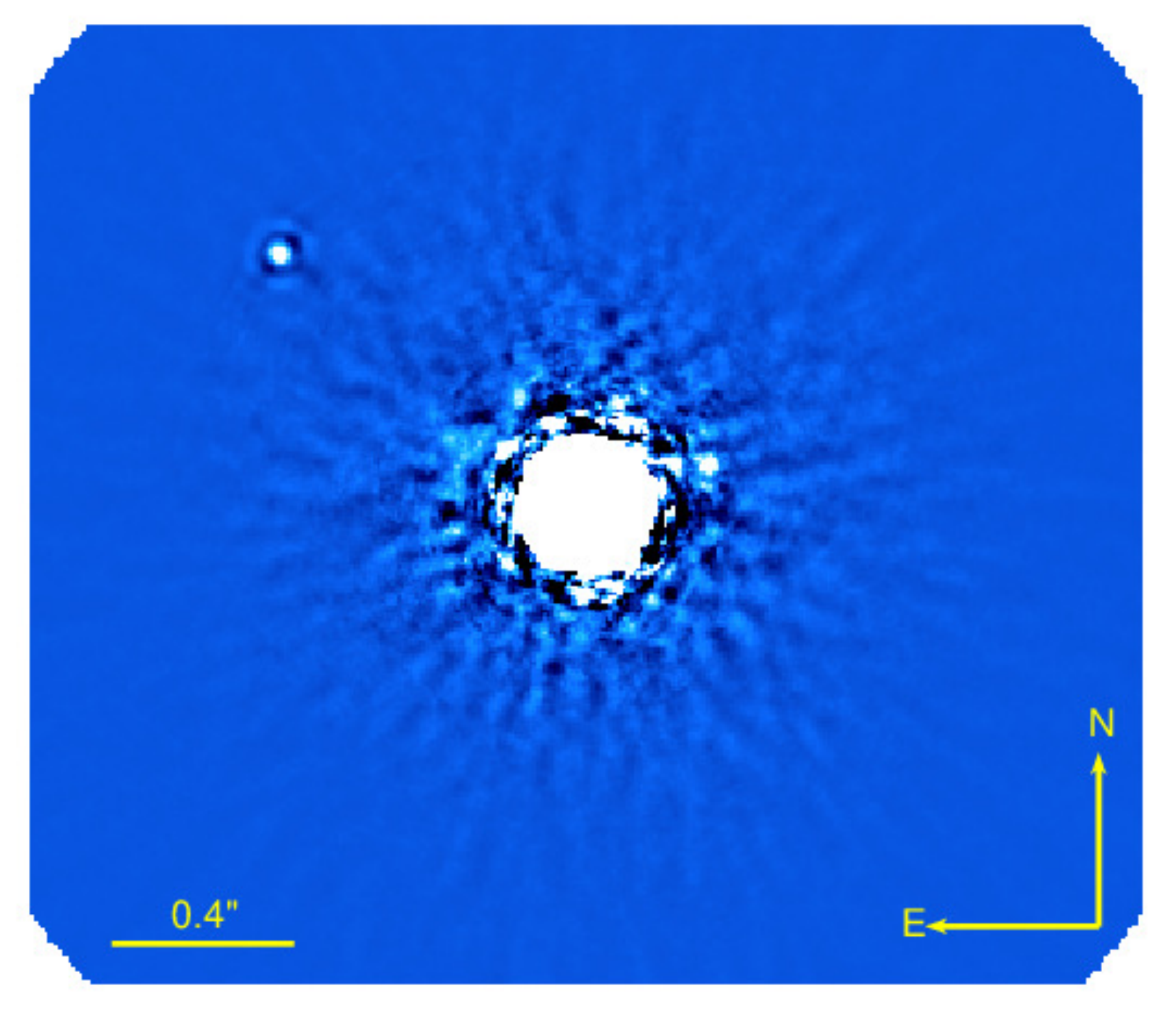}
\end{minipage}
\begin{minipage}{0.5\hsize}
\centering
\includegraphics[scale=0.35]{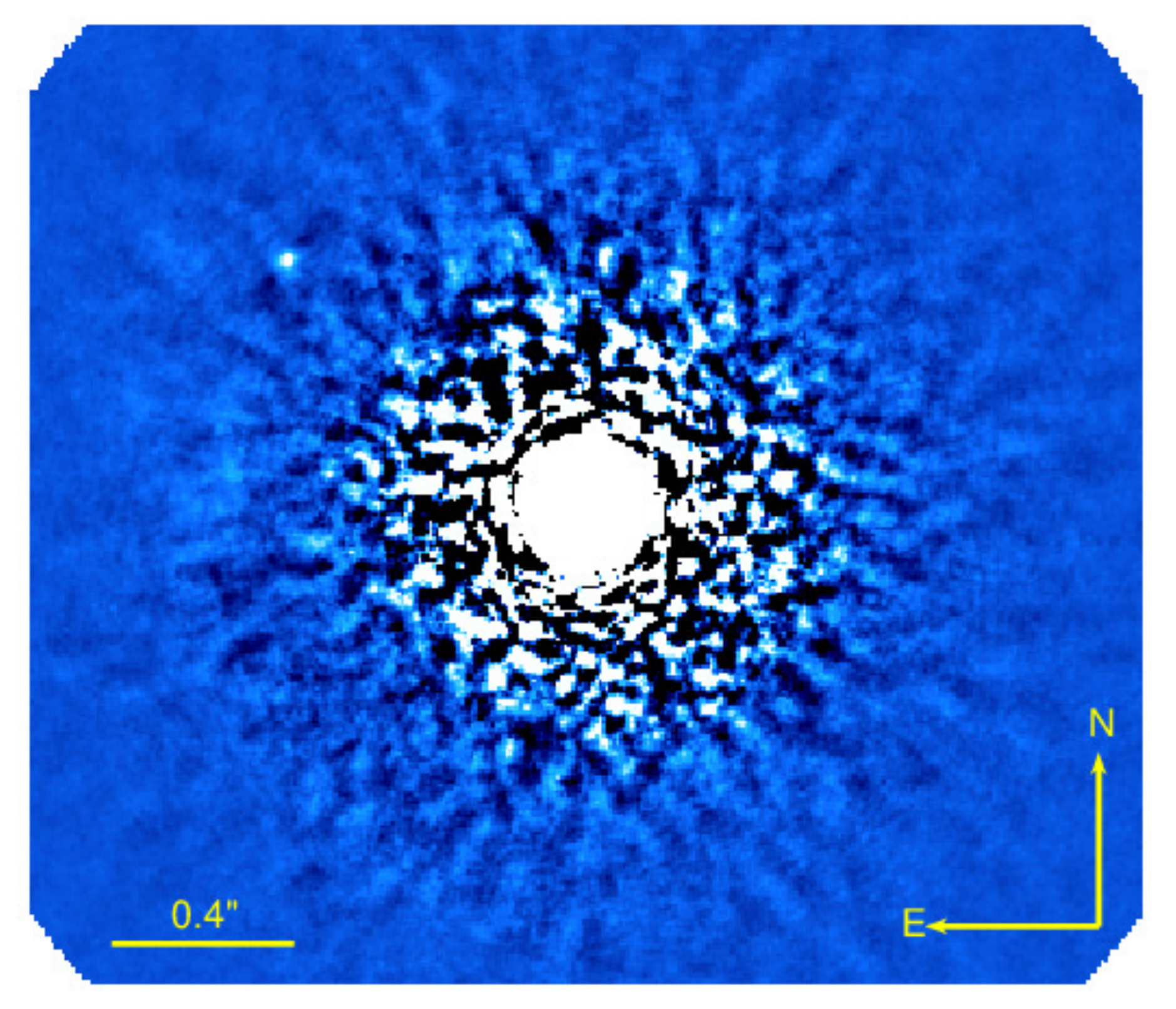}
\end{minipage}
\end{tabular}
\caption{ADI-reduced $\kappa$ And data sets taken by Subaru/SCExAO+HiCIAO in the $H$-band (left) and the $Y$-band (right) in 2016. The central star is masked and the companion is detected in the all images. North is up and east is left.}
\label{SCExAO 2016}
\end{figure*}

\begin{figure*}
\begin{tabular}{cc}
\begin{minipage}{0.5\hsize}
\centering
\includegraphics[scale=0.35]{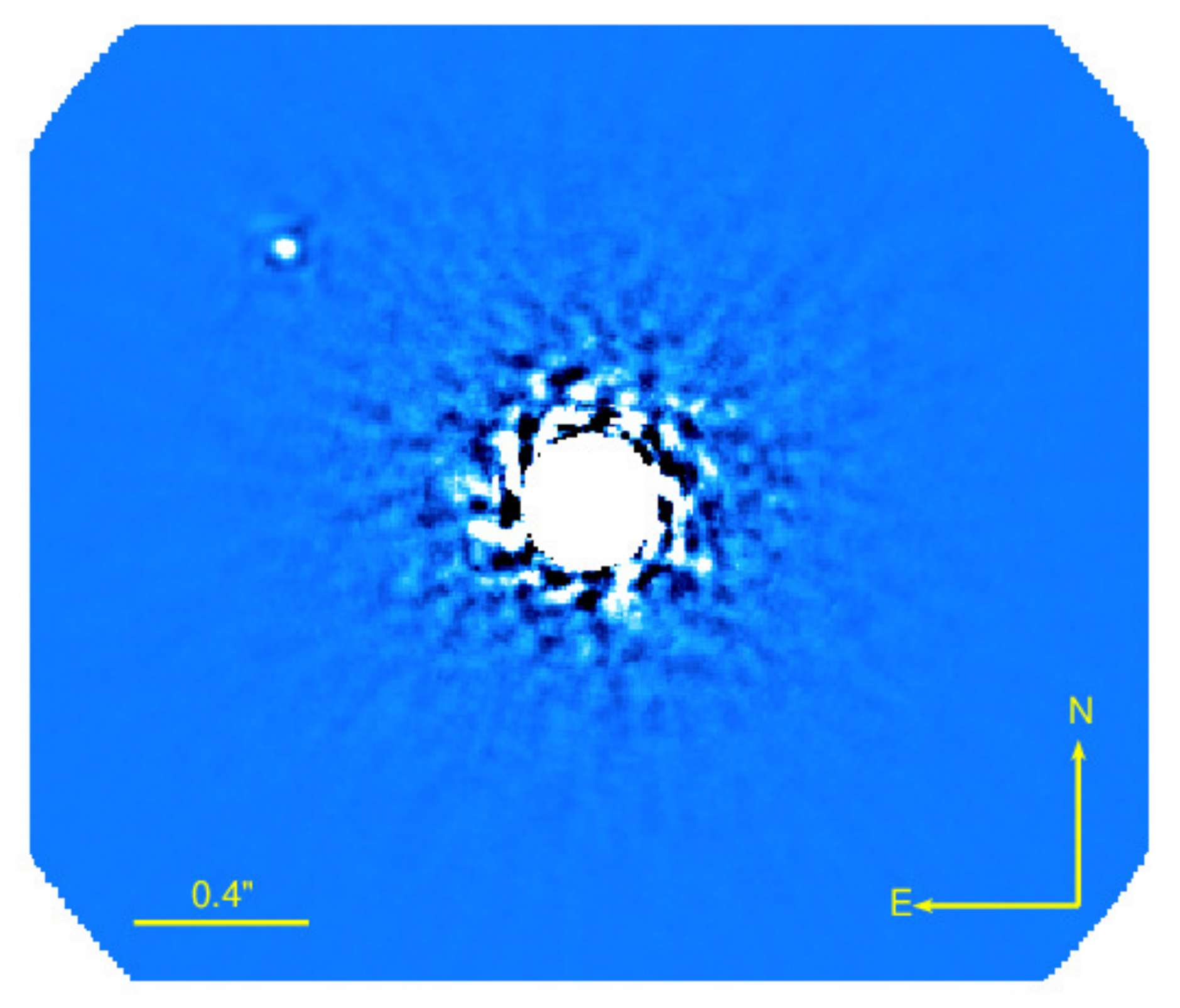}
\end{minipage}
\begin{minipage}{0.5\hsize}
\centering
\includegraphics[scale=0.35]{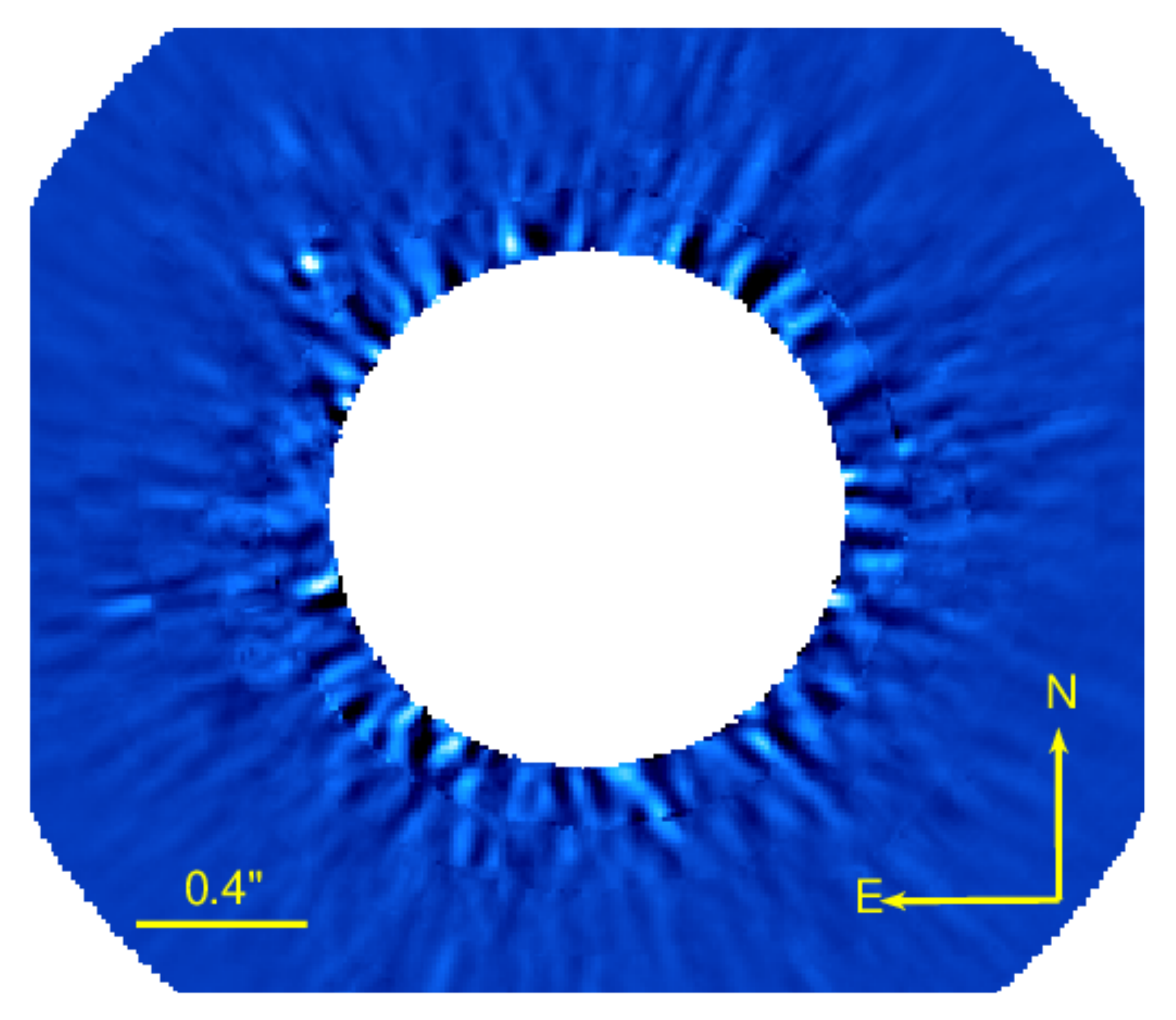}
\end{minipage}
\end{tabular}
\caption{As Figure \ref{SCExAO 2016} for the SCExAO engineering run taken in 2015 (left) and Keck/NIRC2 (right).}
\label{eng and nirc2}
\end{figure*}

Our data reduction detected $\kappa$ And b with signal-to-noise ratios (SNRs) of $\sim$ 10 in the $Y$-band and $\sim$ 130 in the $H$-band (see Figure \ref{SCExAO 2016}) for the 2016's SCExAO+HiCIAO data sets, and SNR$\sim$14 in the Keck/NIRC2 data, respectively.
We also detected $\kappa$ And b with an SNR of $>80$ in the 2015 engineering data (see Figure \ref{eng and nirc2}). 
Compared to \cite{Carson2013} who measured an SNR$\sim$ 20--25 in the $H$-band with Subaru/HiCIAO+AO188, our $H$-band data yielded higher SNR detections.
\cite{Hinkley2013} used Project 1604/Palomar integral field spectroscopy (IFS) to extract $\kappa$ And b's spectrum in $YJH$-bands.   Over the five channels encompassing $Y$ band, the mean ratio of their flux to flux uncertainty is $\sim$ 3, where uncertainties are drawn from the local properties of the noise.  Assuming no contribution from systematic uncertainties and a SNR gain from median-combining channels scaling with the square-root of the number of channels, their band-integrated SNR should be $\sim$ 6.5 or less.   
Thus, our $Y$-band data likely detect $\kappa$ And b at a higher SNR.
The $H$-band detections are comparable in significance to that achieved with high-quality SCExAO+CHARIS data from \citet{Currie2018} due to our data's greater depth and field rotation.

\begin{figure}
\centering
\includegraphics[scale=0.6]{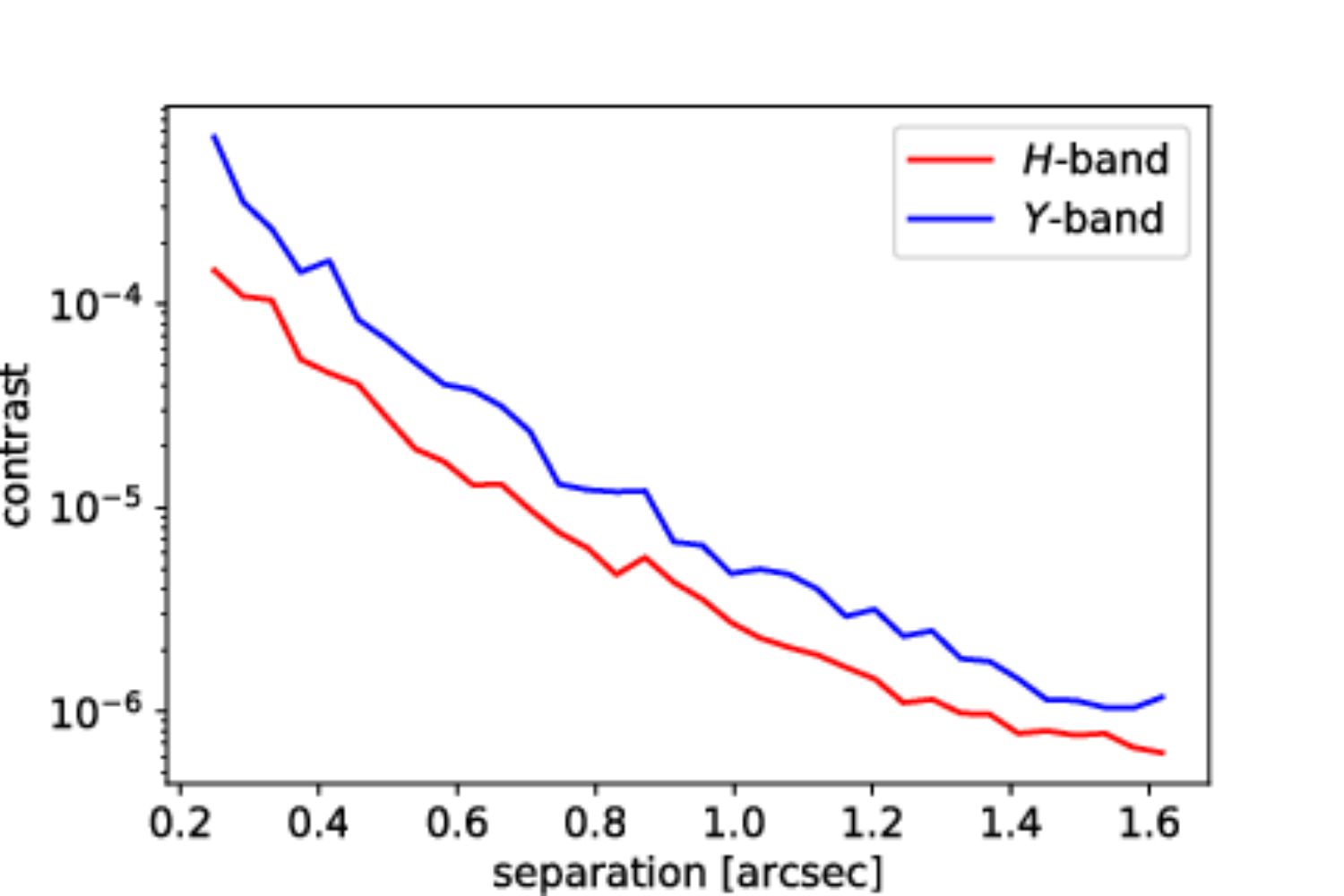}
\caption{5$\sigma$ contrast limits of SCExAO+HiCIAO observations taken in 2016.}
\label{contrast SCExAO+HiCIAO}
\end{figure}

We also calculated contrast limits for $\kappa$ And data sets (see Figure \ref{contrast SCExAO+HiCIAO}).
We convolved the final images, which were normalized with exposure times, and extracted noise profiles from them. Figure \ref{contrast SCExAO+HiCIAO} shows the calculated 5$\sigma$ contrast limits of SCExAO+HiCIAO observations. 
The $H$-band achieved a better contrast level than the $Y$-band observation; 5$\sigma$ contrast limit is 1.5$\times$10$^{-4}$, 2.8$\times$10$^{-5}$, and 2.7$\times$10$^{-6}$ at 0.25$^{\prime\prime}$, 0.5$^{\prime\prime}$, and 1$^{\prime\prime}$, respectively.   At $\rho$ $\sim$ 0\farcs{}3--0\farcs{}75, the planet-to-star contrasts for the SCExAO/CHARIS broadband data in \citet{Currie2018} are about a factor of 2--5 better than those reported here for SCExAO/HiCIAO at $H$ band due to the CHARIS data's better PSF quality and utilization of ADI+SDI for PSF subtraction.  Similarly, the SCExAO/HiCIAO $H$-band contrasts in \citet{Kuhn2018}, which were taken on a different date: 2016 November 12 UT, are typically a factor of 2 deeper, likely due to usage of the vector vortex coronagraph.

\subsection{Photometry and Astrometry} \label{sec: Photometry and Astrometry}

We used aperture photometry for measuring photometry and PSF fitting for estimating FWHM and astrometry in this section.
For absolute photometric calibration, we primarily relied on unsaturated images of other stars obtained through well-calibrated neutral density filters.
As photometric reference of the $Y$-band image to calibrate both $\kappa$ And A and $\kappa$ And b, HIP 118133 \citep[$Y$-band magnitude of 6.60$\pm$0.06 mag; ][]{Pickles2010} was used.
HIP 118133 was observed immediately after kappa And and at a comparable air mass.\footnote{The difference in AO performance between HIP 118133 and $\kappa$ And was insignificant for the purposes of photometric calibration \citep[see also][]{Currie2019b}.}  The implied $Y$-band photometry for the $\kappa$ And A (4.28 $\pm$ 0.09) is consistent with the primary having (near-)zero infrared colors, as expected for a B9V star \citep[e.g.][]{Currie2010,Pecaut2013}.  

We also checked our $H$-band photometric results.
Although unsaturated frames of $\kappa$ And in the $H$-band were taken at both epoch (2015 and 2016), those data used the ND0.1 filter, which was reported to have high uncertainty in its transmission efficiency.
Therefore, we used another set of unsaturated images of HIP 79977, which has $H$-band magnitude of 7.854$\pm$0.03 mag \cite[2MASS;][]{Cutri2003}, for the $H$-band photometric reference. 
In the engineering run, the ND1 filter was used to take one unsaturated frame and we used this image as the photometric reference.

To estimate the throughput correction for $\kappa$ And b needed to compensate for signal loss due to PSF subtraction as well as the astrometric biasing, we injected synthetic companions that are made from an unsaturated PSF of the central star observed through the neutral density filter in each bandpass or (for Keck) with an intensity distribution approximating the star as seen through the partially transmissive coronagraph mask.   
In $H$- band and $K_{\rm s}$- band, we calculated the throughput correction and astrometric biasing over a FWHM-wide area.
In $Y$-band, we adopted a smaller aperture (4.4 pixels or 37 mas), corresponding to most of the PSF core and the apparent PSF size of the real $\kappa$ And b.   
To confirm the reliability of our PSF model at $Y$-band, we verified that the FWHM of the partially-annealed synthetic planet PSF matches that of the real $\kappa$ And b.  The signal throughput in each case is high -- above 80\% for all data sets and $\sim$ 90\% for the Keck/NIRC2 data.


Table \ref{Photometric result} shows our photometric results for the $\kappa$ And system.   Our $H$-band photometry agrees with that derived from SCExAO/CHARIS \citep[$H$ = 15.01$\pm$0.07; ][]{Currie2018} and earlier AO188/HiCIAO photometry from \citet{Bonnefoy:2014dx} ($H$=14.95$\pm$0.13).
Because the photometric uncertainty with our data is higher than with the SCExAO/CHARIS results, 
we use only our $Y$-band result to update photometric parameters of $\kappa$ And b for atmospheric analysis.   The $H$-band data are used for astrometric analysis.
Table \ref{Astrometry Kap And b} summarizes astrometric results of our data sets as well as previous studies\footnote{The 2015 HiCIAO data provided ($\Delta$RA, $\Delta$Dec)=(0.767$\pm$??, 0.638$\pm$??). We have unknown systematic errors due to no distortion correction applied in the SCExAO engineering data. This data set is not presented in Table \ref{Astrometry Kap And b}.}. As mentioned above we calculated astrometric biases when we estimated throughputs by injecting fake sources, which is included in the errors.
The major contributors for the astrometric errors are the intrinsic SNR of the detection and the uncertainty in the centroid position. In case of the Keck data set, we have $0\farcs003$ errors in x \& y position measurement of b and half a pixel uncertainties of in the centroid measurement, which resulted in $0\farcs006$ errors in Table \ref{Astrometry Kap And b}. The centroid was measured by using the PSF seen underneath the partially transmissible coronagraph mask, which gave a better SNR for b than estimating the centroid using the halo outside the mask.
Orbital fitting using these results is described in Section \ref{sec: Astrometry}.

\begin{table}
\centering
\caption{Photometric results of our work}
\begin{tabular}{ccc}\\ \hline\hline
 band& $\kappa$ And A [mag] & $\kappa$ And b [mag] \\ \hline
 $H$& \dots & 15.18$\pm$0.56\footnotemark[1] \\
 $Y$&4.28$\pm$0.09  & 17.04$\pm$0.15 \\ \hline
\end{tabular}
\label{Photometric result}
\footnotetext[1]{Large uncertainty that can be related to unknown offset of the engineering run and different photometric reference in the 2016 data.}
\end{table}

\begin{table}
  \caption{Summary of photometry of $\kappa$ And system}
  \centering
  \begin{tabular}{cccc}\\ \hline\hline
  band & $\kappa$ And A& $\kappa$ And b &  Ref. \\ \hline
  $Y$ [mag] & 4.28$\pm$0.09 & 17.04$\pm$0.15 & \footnotemark[1] \\
  $J$ [mag] & 4.26$\pm$0.04  & 15.84$\pm$0.09 & \footnotemark[2] \\
  $H$ [mag] & 4.31$\pm$0.05 & 15.01$\pm$0.07 & \footnotemark[2] \\
  $K_{\rm s}$ [mag] & 4.32$\pm$0.05 & 14.37$\pm$0.07 & \footnotemark[2] \\
  $L^\prime$ [mag] & 4.32$\pm$0.05 & 13.12$\pm$0.1 & \footnotemark[3], \footnotemark[4] \\
  $NB\_4.05$ [mag] & 4.32$\pm$0.05 & 13.0$\pm$0.2 & \footnotemark[4] \\
  $M^\prime$ [mag] & 4.30$\pm$0.06 & 13.3$\pm$0.3 & \footnotemark[4] \\ \hline
  \end{tabular}
  \label{Photometry summary Kap And}
  \footnotetext[1]{This work} 
  \footnotetext[2]{\cite{Currie2018}} 
  \footnotetext[3]{\cite{Carson2013}}
  \footnotetext[4]{\cite{Bonnefoy:2014dx}}
\end{table}

\begin{table*}
\caption{$\kappa$ And b's relative locations}
\begin{minipage}{\textwidth}
\centering
\scalebox{1}{
\begin{tabular}{ccccc} \\ \hline\hline
Date ($UT$)& instrument & $\Delta$RA [$^{\prime\prime}$] & $\Delta$Dec [$^{\prime\prime}$]& Ref. \\ \hline
   2012-01-01 & Subaru/AO188+HiCIAO &0.884$\pm$0.010 & 0.603$\pm$0.011& \footnotemark[3]  \\
   2012-07-08 & Subaru/AO188+HiCIAO &0.877$\pm$0.007 & 0.592$\pm$ 0.007 & \footnotemark[3] \\ 
   2012-11-03 & Keck/NIRC2 & 0.846$\pm$0.010 & 0.584$\pm$0.010 & \footnotemark[2], \footnotemark[4] \\ 
   2013-08-18 & Keck/NIRC2 & 0.829$\pm$0.010 &0.585$\pm$0.010 &  \footnotemark[2]  \\
   2016-07-18 & Subaru/SCExAO+HiCIAO & 0.734$\pm$0.008 & 0.599$\pm$0.007 & \footnotemark[1] \\
   2017-09-05 & Subaru/SCExAO+CHARIS &0.710$\pm$0.016 & 0.576$\pm$0.012 & \footnotemark[2] \\
   2017-12-09 & Keck/NIRC2 & 0.699$\pm$0.010 & 0.581$\pm$0.010 & \footnotemark[2] \\ 
   2018-11-01 & Keck/NIRC2 & 0.656$\pm$0.006 & 0.580$\pm$0.006 & \footnotemark[1] \\
    \hline
\end{tabular}
}
\label{Astrometry Kap And b}
 \footnotetext[1]{This work}
 \footnotetext[2]{\cite{Currie2018}} 
 \footnotetext[3]{\cite{Carson2013}} 
 \footnotetext[4]{\cite{Bonnefoy:2014dx}}
  \end{minipage}
\end{table*}

\section{Empirical Comparisons to $\kappa$ And b's Photometry and Spectra}
We add $\kappa$ And b's $Y$-band photometry to CHARIS $JHK$ spectra to provide a new empirical context for the companion's near-infrared properties.   Previous empirical spectral analysis from \citet{Currie2018} using spectral templates and a homogeneously reduced library of substellar object spectra, pointed towards $\kappa$ And b being an L0-L1 low surface gravity object consistent with a young, planet-mass companion.   Our new data extend the available wavelength baseline for $\kappa$ And b data.   We compare $\kappa$ And b's broadband photometry to field and low gravity objects using a larger set of empirical substellar object spectra. 
\subsection{Near-Infrared Colors}
\label{sec: Comparison with Spectral Templates}

\begin{figure}
\centering
\includegraphics[width=1.0\columnwidth]{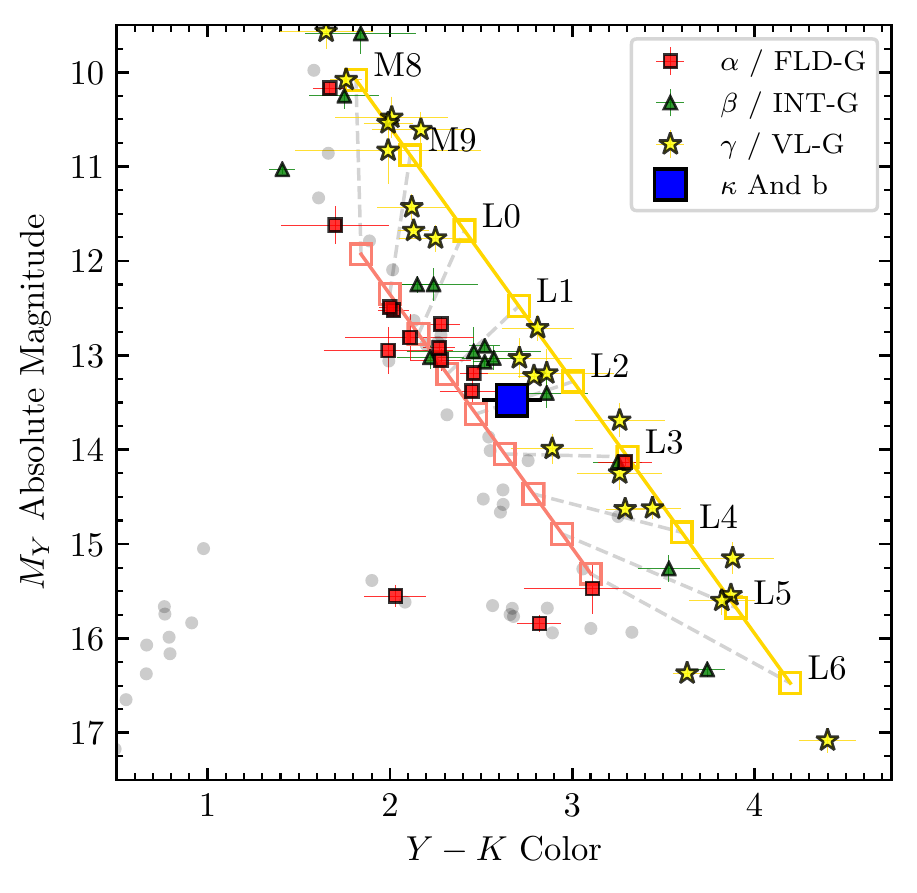}
\caption{Near-infrared color-magnitude diagram showing $\kappa$ And b (blue square) relative to other substellar objects with trigonometric parallax measurements from \citet{Liu:2016co}. Objects without a literature gravity classification are denoted by small gray circles. Linear fits to the absolute magnitude and colors of field gravity (red open squares) and low gravity (yellow open squares) are also shown. $\kappa$ And b appears somewhat redder than field-gravity objects with a similar $Y$-band absolute magnitude{\bf, but not at a significant level}.}
\label{fig:cmd}
\end{figure}

\begin{figure}
\centering
\includegraphics[width=1.0\columnwidth]{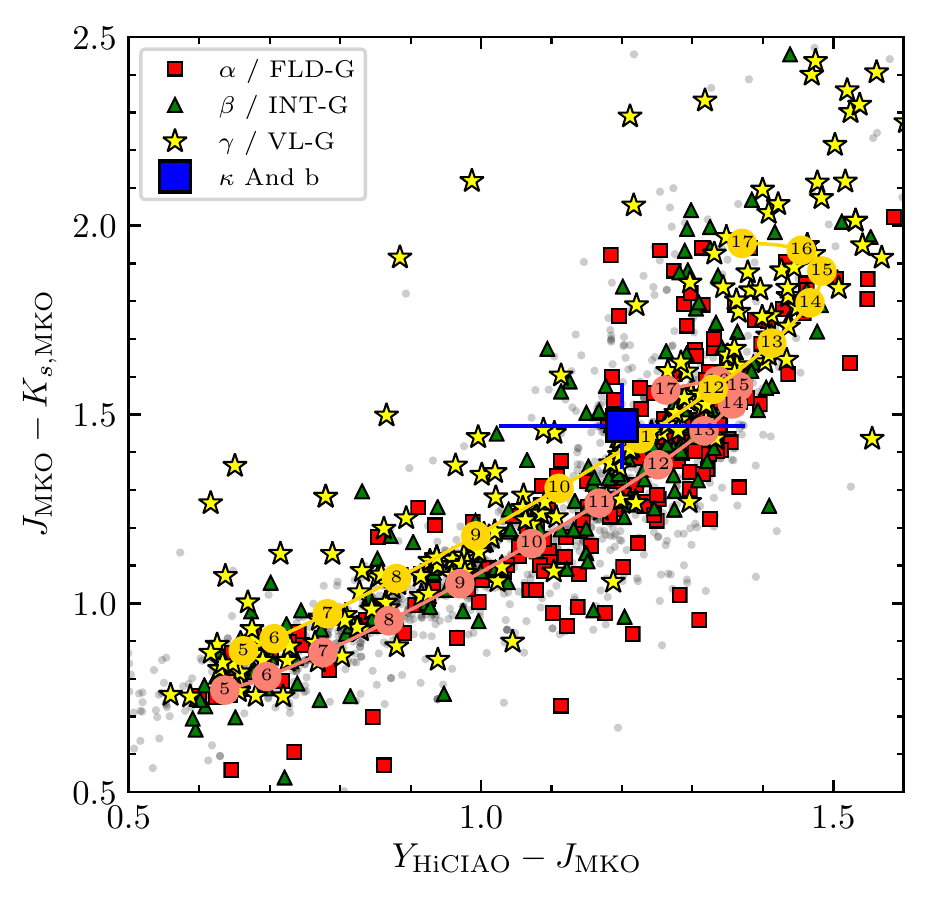}
\caption{Near-infrared color-color diagram magnitude diagram showing $\kappa$ And b (blue square with error bars) compared to $Y-J$ and $J-K$ colors for objects in our spectral library. Third-order polynomial fits to the color as a function of spectral type are plotted as red (field gravity) and yellow (very low gravity) lines, with M0=0, L0=10, etc. Red squares, green triangles, yellow stars, and blue diamonds denote objects with field, intermediate gravity/$\beta$, very-low gravity/$\gamma$, and $\delta$ gravities, respectively.   Gray dots denote dwarfs without gravity classifications.}
\label{fig:colcol}
\end{figure}

\begin{figure}
\centering
\includegraphics[width=1.0\columnwidth]{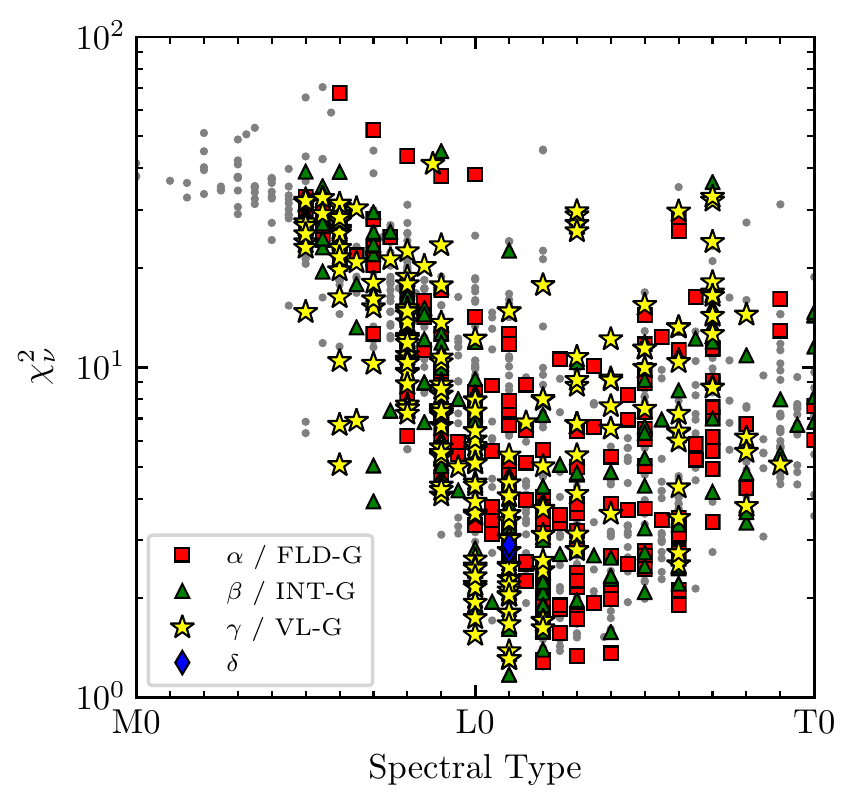}
\caption{Goodness of fit as a function of spectral types for the objects within the spectral library compared to the near-infrared photometry and low-resolution spectroscopy of $\kappa$~And~b. Comparison objects with previously published gravity classifications in the literature are highlighted.  Our analysis shows that some objects with low $\chi^{2}$ that are either unclassified or previously classified as field dwarfs/intermediate gravity dwarfs may in fact be low-gravity objects (see text).}
\label{fig:chi2_spectraltype}
\end{figure}

\begin{figure*}
\centering
\includegraphics[width=1.0\textwidth]{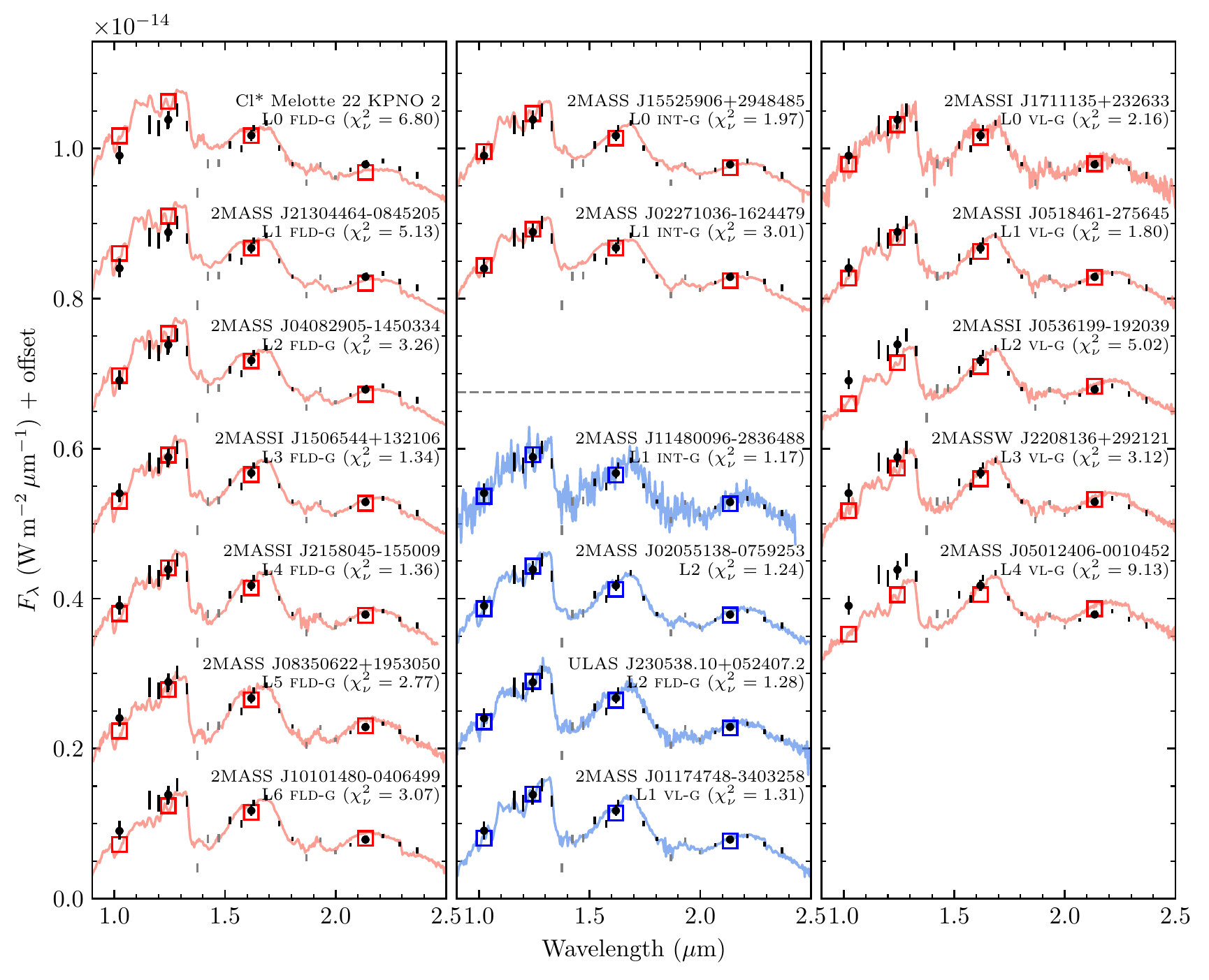}
\caption{The near-infrared SED of $\kappa$~And~b (black points) compared to the early L-type near-infrared standards proposed by \citet{Cruz:2018ih} for field (L0--L6$\alpha$, left column), intermediate (L0--L1$\beta$, middle column), and very-low (L0--L4$\gamma$, right column) surface gravities (red). CHARIS spectral channels within the water absorption bands were not included in the fit (gray points). Four of the best fit objects within the complete library are also plotted in the central column (blue).  Their previously published gravity classifications are given; our analysis revises some of them to lower gravity classes. Spectra are from \citet{Burgasser:2006jj, Kirkpatrick:2010dc, BardalezGagliuffi:2014fl, Burgasser:2007fl, Burgasser:2010df, Chiu:2006jd, Reid:2006da, Allers2013, Burgasser:2008cj, Gagne2015, Kellogg:2017kh, Cruz:2018ih, Filippazzo:2015dv}.}
\label{fig:bestfit_library}
\end{figure*}

We first investigated a color-magnitude diagram of $\kappa$ And b by comparing it to other low-mass objects with precise parallaxes and various gravities reported in \citet{Liu:2016co}.   The \citet{Liu:2016co} sample includes 67 MLT dwarfs with new, precise parallaxes and another 35 with literature parallaxes and near-infrared photometry.  Drawing from the \citet{Liu:2016co} polynomial fits for absolute magnitudes vs. spectral for different gravity classes, we constructed linear fits to magnitudes and colors in Y/Y-K space.

Figure \ref{fig:cmd} shows how $\kappa$ And b's color magnitude diagram position fits within the context of other substellar objects.   The companion appears redder than a typical field-gravity L object (red), in between these colors and those for typical low-gravity L object (yellow) at its $Y$-band luminosity.  Moreover, its location appears on the locus (grey dashed line) connecting L2 field and low surface gravity objects.   
The uncertainty of the $Y-K$ color of $\kappa$ And b and the amplitude of the scatter of objects about the polynomial fits from \citet{Liu:2016co} preclude us from excluding a high or low gravity scenario at a significant level using only the $Y$ and $K$ band luminosities.

Second, we use a large sample of substellar objects with different spectral types and gravity classifications to provide a context for $\kappa$ And b's near-infrared colors.  We compiled a library of 2,011 M-, L-, and T-dwarf spectra drawn from the SpeX Prism library\footnote{\url{http://pono.ucsd.edu/~adam/browndwarfs/spexprism/}} \citep{Burgasser2014}, the IRTF Spectral Library\footnote{\url{ http://irtfweb.ifa.hawaii.edu/~spex/IRTF_Spectral_Library/}} \citep{Cushing2005}, the Montreal Spectral Library\footnote{\url{ https://jgagneastro.wordpress.com/the-montreal-spectral-library/}} (e.g., \citealp{Gagne2015,Robert2016}), and the sample of young ultracool dwarfs presented in \cite{Allers2013}. We do not incorporate the library of young, low-gravity objects presented in \citet{Bonnefoy:2014dh} and used by \citet{Currie2018} in their analysis of $\kappa$~And~b as the SINFONI spectra do not extend into the $Y$-band and thus cannot be compared to the new photometry presented in this work. The spectral types were obtained from a number of literature sources, and are given for a number of sources highlighted in the remainder of this section. We preferentially used the near-infrared spectral type if both an optical and near-infrared classification were available. Gravity classifications for a subset of the objects were also obtained from the literature, using either of the schemes outlined by \citet{Kirkpatrick2005,Kirkpatrick2006,Cruz2009} ($\alpha$, $\beta$, $\gamma$, $\delta$ in descending order of surface gravity), or \citet{Allers2013} ({\sc fld-g}, {\sc int-g}, {\sc vl-g}, similarly). Both of these classifications schemes share three categories; surface gravity indicators consistent with those observed in old field dwarfs ($\alpha$, {\sc fld-g}), and intermediate surface gravity ($\beta$, {\sc int-g}), and a very low surface gravity observed for substellar objects in nearby, young moving groups ($\gamma$, {\sc vl-g}). The fourth classification $\delta$ was defined by \cite{Kirkpatrick2005} for objects that exhibit stronger gravity-sensitive features than seen for those classified as $\gamma$/{\sc vl-g}.

We computed synthetic $Y_{\rm HiCIAO}$, $J_{\rm MKO}$, $H_{\rm MKO}$, and $K_{s, \rm MKO}$ photometry for the library by convolving the spectra with appropriate filter response curves given in Figure~\ref{Y-band transmission} and in \citet{Tokunaga2002}. Figure \ref{fig:colcol} compares $\kappa$ And b's $Y-J$ and $J-K$ colors to library objects with different gravity classifications.   The main locus of library colors extends from $Y-J$/$J-K$ $\sim$ 0.6/0.8 to 1.3/1.5 for M5 dwarfs to L3 dwarfs.   Young objects with intermediate or (very-)low gravities appear systematically redder in $J-K$, as expected from previous studies \citep{Liu:2016co}. $\kappa$ And b's position lies between typical L0 and L2 colors, above positions for most field objects and overlapping with younger, lower gravity objects.

\subsection{Joint $Y$-band photometry and CHARIS $JHK$ Spectral Comparisons}

To assess the overall best-fitting objects among the libraries, we fit $\kappa$ And b's $Y$ band photometry and CHARIS spectra.   Library spectra were convolved and interpolated to CHARIS's wavelengths and spectral resolution, assuming a constant resolution of $R=20$ across the full spectrum. We removed 20 library spectra that did not have wavelength coverage spanning the $Y$- through $K$-bands. A small subset of the library had H-band spectra that were truncated at $\sim$1.75\micron, shorter than the reddest $H$-band channel in the CHARIS spectrum at 1.8\micron. For these 135 spectra, we excluded this CHARIS channel from the fit and reduced the number of degrees of freedom by one when calculating $\chi^2_{\nu}$. 

We computed the goodness of fit for each object by calculating $\chi^2_{\rm spec}$ from a comparison of the $\kappa$~And~b spectrum to the smoothed library spectra using the correlation matrix given in \citet{Currie2018}, and $\chi^2_{\rm phot}$ from a comparison of the near-infrared photometry of $\kappa$~And~b to the synthetic photometry of the objects within the library. As we were primarily interested in comparing the spectral morphology of $\kappa$~And~b to the objects within the library, we computed the scaling factor to apply to the library spectrum and photometry that minimized $\chi^2 = \chi^2_{\rm spec} + \chi^2_{\rm phot}$. We did not incorporate the library spectra measurement uncertainty; these were typically negligible when convolved to CHARIS's resolution.

Figure~\ref{fig:chi2_spectraltype} displays the $\chi^2_{\nu}$ distribution for M0--T0 objects in the library.  Early L-type objects show a clear minimum, consistent with analyses presented in \citet{Bonnefoy:2014dx} and \cite{Currie2018}. The exact location of the minimum differs for field and low-gravity objects; at L1 for $\gamma$/{\sc vl-g} objects and at L2--L3 for $\alpha$/{\sc fld} objects, a consequence of the redder near-infrared colors of low-gravity objects compared to field objects of the same spectral type (e.g. \citealp{Liu:2016co}, Fig. 15). 
This effect is also seen when comparing $\kappa$~And~b to the L-type standards proposed by \citet{Cruz:2018ih}, shown in Figure~\ref{fig:bestfit_library}, where the best-fit low-gravity standard is L1 ($\chi^2_{\nu}=1.8$) and later spectral types (L3--L4) fit far worse, while the best-fit field gravity standards are L2--L3 and earlier spectral types (e.g. L0) fit far more poorly. This trend is consistent with that seen for synthetic spectral templates (composites of individual spectral standards for a given spectral type/gravity class, \citealp{Cruz:2018ih}) in \citet{Currie2018}: they found that the best-fit low gravity template (L0$_{\gamma}$, $\chi^{2}_{\nu}$ $\sim$ 1.26) is three subtypes earlier than the best-fit field gravity template (L3, $\chi^{2}_{\nu}$ $\sim$ 1.51).

Of the objects within the complete library, the best fit was \object{2MASS J11480096-2836488} ($\chi^2_{\nu}=1.2$), previously classified an L1 intermediate-gravity member of the 10\,Myr \citep{Bell:2015gw} TWA moving group  \citep{Gagne2015,Gagne:2018jj} and an isochronal mass of $\sim$8\,M$_{\rm Jup}$ \citep{Gagne2015}. While the signal-to-noise ratio of the spectrum for this object is lower than the typical library spectrum, the uncertainties are comparable to those of the spectrum of $\kappa$~And~b when degraded to the same resolution. Good fits were also found to \object{2MASS J01174748-3403258} ($\chi^2_{\nu}=1.3$; previously classified as L1 $\gamma$) and to \object{2MASS J02055138-0759253} and \object{ULAS J230538.10+052407.2} ($\chi^2_{\nu}=1.2$ and 1.3), which previously were unclassified or classified as being field gravity L2 dwarfs. In total, 36 objects have a $\chi^2_{\nu}<1.7$ (95\% confidence level) with the following previous classifications: 1 L0 ({\sc vl-g}), 5 L1 (2 {\sc int-g}, 3 {\sc vl-g}), 22 L2 (11 without classification, 8 {\sc fld-g}, 2 {\sc int-g}, 1 {\sc vl-g}), 4 L3 (3 without classification 1 {\sc fld-g}), and 4 L4 (2 without classification, 1 {\sc fld-g}, and 1 {\sc int-g}). For reference, the complete library contains 656 objects between L0--L4; 381 without classification, 112 {\sc fld-g}, 80 {\sc int-g}, and 81 {\sc vl-g}.

To further investigate the nature of the four best-fit objects, we separately estimated spectral types using and derived gravity classifications following the spectral index-based methods in \citet{Allers2013}: i.e. the H$_{2}$0, H$_{2}$0-1, H$_{2}$0-2, and H$_{2}$0-D indices for spectral typing and Fe$_{\rm z}$, VO, KI$_{\rm J}$, and $H_{\rm cont}$ for gravity scoring.   We nominally box-car smooth the spectrum using a window size of 3 spectral channels and explore results obtained with different windows.   Our analysis recovers the previous classification for \object{2MASS J01174748-3403258} (L1 $\gamma$).   However, it favors reclassifying \object{2MASS J02055138-0759253} and \object{ULAS J230538.10+052407.2} as L2 $\beta$ objects (gravity scores 1111 and 1120), respectively; Banyan-$\Sigma$ suggests that \object{2MASS J02055138-0759253}'s kinematics may be consistent with membership in the 40 Myr-old Columba association, depending on its parallax.    Given the noisiness of \object{2MASS J01174748-3403258}'s spectrum, we cannot derive a gravity score from Fe$_{\rm z}$, VO, and KI$_{\rm J}$.  However, its $H_{\rm cont}$ index (1.05 $\pm$ 0.05) suggests a low gravity and possible reclassification to L1 $\gamma$.  It is likely that the other well-fitting objects previously given a field classification or no classification at all are in fact low-gravity objects.

To investigate the constraining power of our new $Y$-band photometry, we compared the $\chi^2$ for each object with and without this measurement. For objects between L0 and L1 we typically find a larger $\Delta\chi^2$ for field-gravity objects (median $\Delta\chi^2$ of 4.1 compared to 1.2), indicating that the $Y$-band photometry is more consistent with that of a low-gravity object over this range of spectral types. For later spectral types this is reversed, with $\Delta\chi^2$ typically being larger for low-gravity objects between L2 and L5 (median $\Delta\chi^2$ of 6.9 compared to 0.7). This is a consequence of the red color of low-gravity objects; an object with a given $Y$-band flux (or $Y-J$ color) either has lower gravity and an earlier spectral type, or a higher gravity and a later spectral type.

\begin{figure}
\centering
\includegraphics[width=1.0\columnwidth]{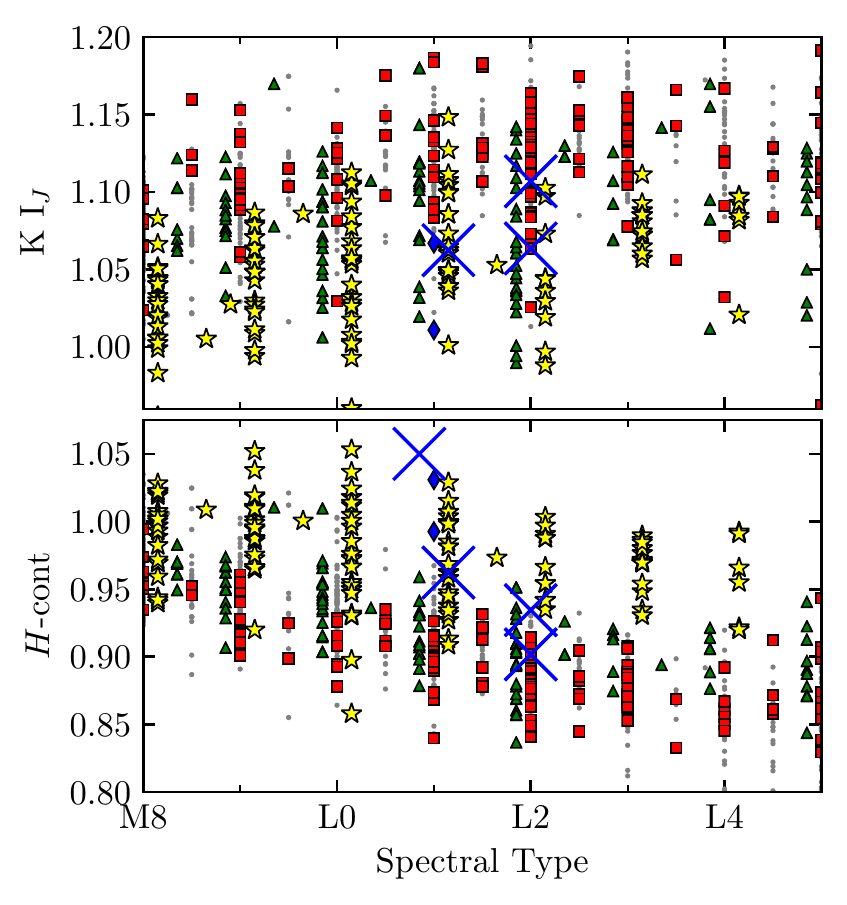}
\caption{Surface gravity indicators from \citet{Allers2013} as a function of spectral type for the objects within the spectral library. Symbols as in Figure~\ref{fig:chi2_spectraltype}, with the four best fit objects to the spectrum and photometry of $\kappa$~And~b highlighted with blue crosses (the spectrum of 2MASS J1148 is too noisy for a reliable estimate of its K I$_{J}$ index). The spectral types of intermediate and low gravity objects have been displaced slightly ($\pm0.15$ subtypes) for clarity.}
\label{fig:grav_indicators}
\end{figure}

Preference for a low surface gravity for $\kappa$~And~b can also be inferred using the gravity-sensitive spectral indices defined by \citet{Allers2013}. While these indices cannot be computed directly given the low resolution of the spectrum, they can be computed for the objects within the library with the most similar spectra to $\kappa$~And~b. Two of these indices are plotted in Figure~\ref{fig:grav_indicators}, showing that the best fit objects are more consistent with the population of low-gravity objects and (some) intermediate gravity objects than the median of the field-gravity sequence.


\section{Comparison with Model Atmospheres}
\label{sec: Comparison with Model Atmospheres}
\begin{table*}
\caption{Summary of Atmosphere Models}
\label{tbl:models}
\hspace{-0.8cm}
\begin{tabular}{lllcccc|cccc}
\hline\hline
\multicolumn{7}{c|}{Model Properties} & \multicolumn{4}{c}{Best fit}\\
\hline
Name &
Ref. &
Special Remark &
$T_{\rm eff}$ &
$\log g$ &
$\Delta T_{\rm eff}$ &
$\Delta \log g$ &
$T_{\rm eff}$ &
$\log g$ &
$R$ &
$\chi^2_{\nu}$ \\
& & & (K) & [dex] & (K) & [dex] & (K) & [dex] & ($R_{\rm Jup}$) & \\
\hline
\multicolumn{10}{c}{{\it Clear models}}\\
{\sc AMES-Cond} & \footnotemark[1] & \nodata & 1000--2400 & 2.5--6.0 & 100 & 0.5 & 2400 & 4.0 & 0.74 & 29.7\\
{\sc BT-Cond}   & \footnotemark[2] & \nodata & 1000--2200 & 4.0--5.5 & 100 & 0.5 & 2200 & 4.0 & 0.85 & 20.4\\
Burrows    & \footnotemark[3] & \nodata & 1000--2000 & 4.5--5.5 & 100 & 0.5 & 2000 & 4.5 & 0.90 & 53.9\\
\multicolumn{10}{c}{{\it Cloudy models}}\\
{\sc AMES-Dusty} & \footnotemark[1] & \nodata & 1000--2500 & 3.5--6.0 & 100 & 0.5 & 1800 & 5.0 & 1.19 & 3.62\\
{\sc BT-Dusty}   & \footnotemark[2] & \nodata & 1000--2400 & 4.5--5.5 & 100 & 0.5 & 1800 & 4.5 & 1.64 & 1.81\\
{\sc BT-Settl}      & \footnotemark[2] & \citet{Asplund:2009eu} abundances & 1000--2400 & 3.0--5.5 & 100 & 0.5 & 1900 & 4.5 & 1.23 & 2.80\\
{\sc BT-Settl}      & \footnotemark[2] & \citet{Caffau:2011ik}  abundances & 1000--2400 & 3.5--5.5 & 50 & 0.5 & 1800 & 5.0 & 1.34 & 1.70\\
{\sc BT-Settl}-2015 & \footnotemark[2] & \nodata        & 1200--2400 & 3.0--5.5 & 50 & 0.5 & 1750 & 5.5 & 1.37 & 3.49\\
{\sc BT-Settl}-bc   & \footnotemark[2] & \nodata        & 1100--2400 & 3.0--5.5 & 100 & 0.5 & 1800 & 4.0 & 1.30 & 2.99\\
{\sc Drift-Phoenix} & \footnotemark[4] & \nodata & 1000--2400 & 3.0--6.0 & 100 & 0.5 & 1700 & 4.0 & 1.57 & 1.66\\
{\tt Burrows} & \footnotemark[3] & Nominal cloud model, 100\micron\ modal size (E100) & 1000--2000 & 4.5--5.5 & 50 & 0.1 & 1800 & 4.6 & 1.25 & 7.08\\
Burrows         & \footnotemark[5] & Thick clouds, 4\micron\ modal size (A4)  & 1800--2200 & 3.5--4.0 & 25--100 & 0.25 & 1900 & 4.0 & 1.23 & 6.39\\
Burrows         & \footnotemark[5] & Thick clouds, 10\micron\ modal size (A10) & 1800--2200 & 3.6--4.0 & 100 & 0.1 & 2000 & 4.0 & 1.09 & 3.24\\
\hline
\end{tabular}
\footnotetext[1]{\citet{Allard:2001fh}}
\footnotetext[2]{\citet{Allard:2012fp}}
\footnotetext[3]{\citet{Burrows:2006ia}}
\footnotetext[4]{\citet{Witte:2011kn}}
\footnotetext[5]{\citet{Currie2014a}}
\end{table*}
The CHARIS near-infrared spectrum from \citet{Currie2018}, the $Y$-band photometry presented in this work, and the literature photometry spanning 1.2--4.7\micron\ (Table \ref{Photometry summary Kap And}) were fit to a number of models of substellar atmospheres. These model grids can be broadly categorized into those that incorporate a prescription for the formation of clouds within the photosphere, and those that enforce a clear photosphere over the full range of effective temperatures and surface gravities. The first group contains the {\sc AMES-Cond} \citep{Allard:2001fh}, {\sc BT-Cond} \citep{Allard:2012fp}, and \cite{Burrows:2006ia} model grids. The {\sc AMES-Cond} and {\sc BT-Cond} grids both use the same {\tt PHOENIX} atmosphere code \citep{Hauschildt:1992ff}, but different molecular line lists (\citealp{Partridge:1997kh} and \citealp{Barber:2006dm}, respectively). These two grids ignored dust opacity entirely in order to simulate the immediate sedimentation of dust into the lower atmosphere leading to a clear photosphere. The \citet{Burrows:2006ia} clear atmosphere grid was created using the {\tt TLUSTY} atmosphere code \citep{Hubeny:1995fw} similarly ignoring opacity from condensates within the photosphere.

The second group contains a number of different treatments for photospheric clouds. The {\sc AMES-Dusty} \citep{Allard:2001fh} and {\sc BT-Dusty} \citep{Allard:2012fp} grids were created using the same atmospheric code and line lists as the clear photosphere models described previously, but instead including dust opacity in the calculation of the emergent spectra and neglecting gravitational sedimentation entirely. The various {\sc BT-Settl} grids \citep{Allard:2012fp} were also calculated with the same code, but with a revised treatment for dust sedimentation to better model the L/T transition from cloudy to clear photospheres. The {\sc Drift-Phoenix} grid \citep{Witte:2011kn} used the same {\tt PHOENIX} code, but a completely revised treatment for the formation and evolution of photospheric clouds that reproduces the observed SED of young, low-gravity objects (e.g., \citealp{Patience:2012cx,Lachapelle:2015cx}). The \citet{Burrows:2006ia} models simulate clouds of a variety of condensates as extending between the scale heights set by the most and least refractory condensates, with an exponential decay above and below. The extent of the clouds and the size distribution of particles within the clouds are free parameters within the model. Here we compare to the fiducial cloud model used in \citet{Burrows:2006ia} that has a model particle size of 100\micron, and also to the thick cloud models with smaller modal particle sizes (4\micron\ and 10\micron) used in \citet{Currie2014a}. A summary of the various atmosphere model grids, and their coverage and resolution in  ($T_{\rm eff}$, $\log g$) space, is given in Table~\ref{tbl:models}. 

\begin{figure}
\centering
\includegraphics[width=1.0\columnwidth]{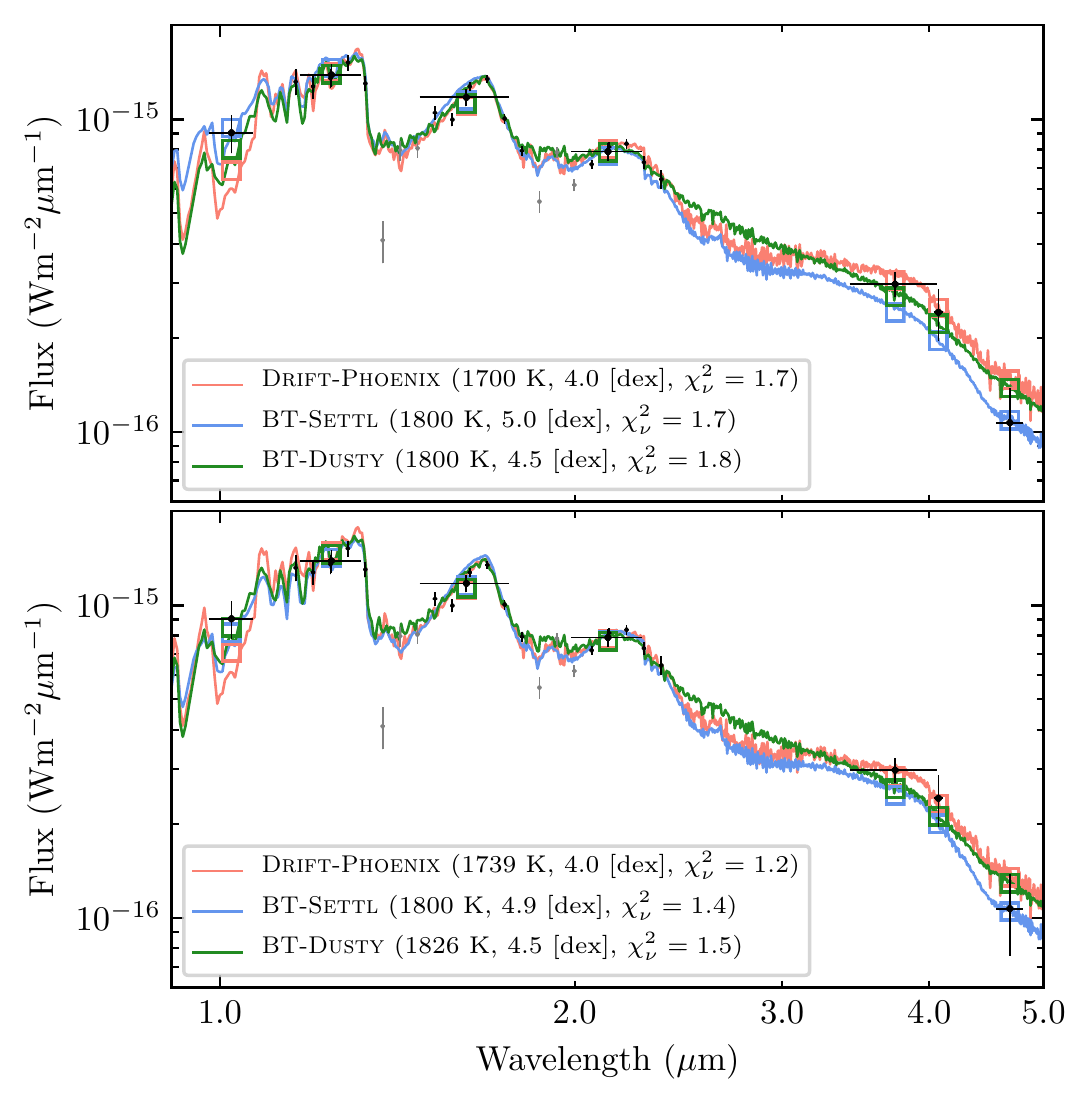}
\caption{The best fit model atmosphere within the {\sc Drift-Phoenix} (red), {\sc BT-Settl} (blue), and {\sc BT-Dusty} (green) to the observed SED of $\kappa$~And~b without interpolation (top) and with interpolation between the grid points of the models. The spectrophotometry of $\kappa$~And~b is overplotted (black), with low-SNR channels of the CHARIS spectrum excluded from the fit shown in gray.}

\label{fig:model_plot}
\end{figure}

\begin{figure}
\centering
\includegraphics[width=1.0\columnwidth]{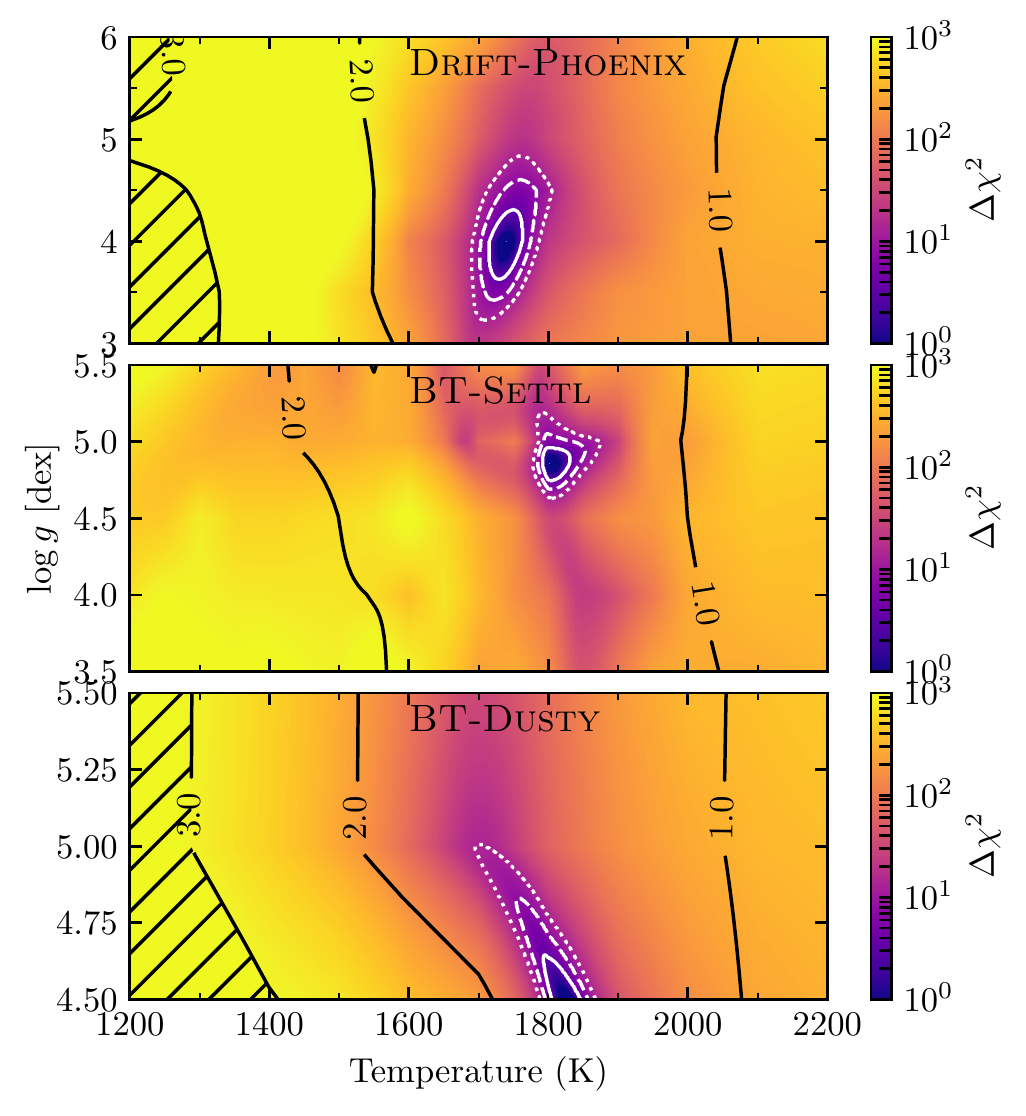}
\caption{$\Delta\chi^2$ surface for the {\sc Drift-Phoenix} (top), {\sc BT-Settl} (middle), and {\sc BT-Dusty} (bottom) grids calculated using the interpolated version of each grid. Black contours denote the radius required to minimize the $\chi^2$; hatched region requires an non-physical radius of $>3$\,$R_{\rm Jup}$. White contours denote 1, 2, and 3\,$\sigma$ credible regions, computed from the $\Delta\chi^2$.}
\label{fig:model_chi2}
\end{figure}

The model atmospheres were fit to the observed photometry and spectroscopy of $\kappa$~And~b using a similar procedure as for the empirical comparison described in Section~\ref{sec: Comparison with Spectral Templates}, including the thermal infrared measurements given in Table~\ref{Photometry summary Kap And}. We applied a limit on the value of the dilution factor ($r^2/d^2$) such that the radius of the companion was between 0.5--3.0\,R$_{\rm Jup}$, encompassing the range of radii predicted for young substellar companions (e.g., \citealp{Fortney:2008ez}). As with \citet{Currie2018}, we assume a distance of $d=50.0$\,pc \citep[Table 1;][]{GaiaDR2-2018}. The best fit model, and corresponding $\chi^2$, within each grid is given in Table~\ref{tbl:models}.

Cloudy models are preferred by a significant margin, although the quality of the fit varies between each grid. Of all the models tested, the best fit was the 1700\,K, $\log g=4.0$\,[dex] model within the {\sc Drift-Phoenix} grid. This model is plotted against the SED of $\kappa$~And~b in Figure~\ref{fig:model_plot}, alongside the two other best-fitting models from the {\sc BT-Settl} and {\sc BT-Dusty} grids. The temperatures of these models are consistent with the spectral type determined previously.   The Burrows model fitting results favor thick clouds and a modal dust size somewhere betwen 4 $\mu m$ and 100 $\mu m$.

Two of the grids displayed in Figure~\ref{fig:model_plot} -- {\sc Drift-Phoenix} and {\sc BT-Dusty} -- suggest a low surface gravity consistent with our empirical comparisons.   The {\sc BT-Dusty} grid only covers a limited range of $\log g$ and the best fit was found at the grid boundary of $\log g = 4.5$\,[dex].  Thus, it is likely that a {\sc BT-Dusty} model grid covering a wider range in gravity (e.g. log(g) = 3--5) would result in a surface gravity approaching that found for {\sc Drift-Phoenix} ($\log g = 4.0$\,[dex]). The best fit model within the the third grid ({\sc BT-Settl}) has a similar goodness of fit but a higher surface gravity ($\log g =5.0$\,[dex]). The range of best fit surface gravities for the three model grids is a reasonable proxy for the model uncertainty, demonstrating both how differences in assumptions regarding cloud properties and extent can affect derived bulk properties, and that the surface gravity of $\kappa$~And~b cannot be conclusively derived from low-resolution spectroscopy and photometry used in this study.

\begin{figure}
    \centering
    \includegraphics[scale=0.7]{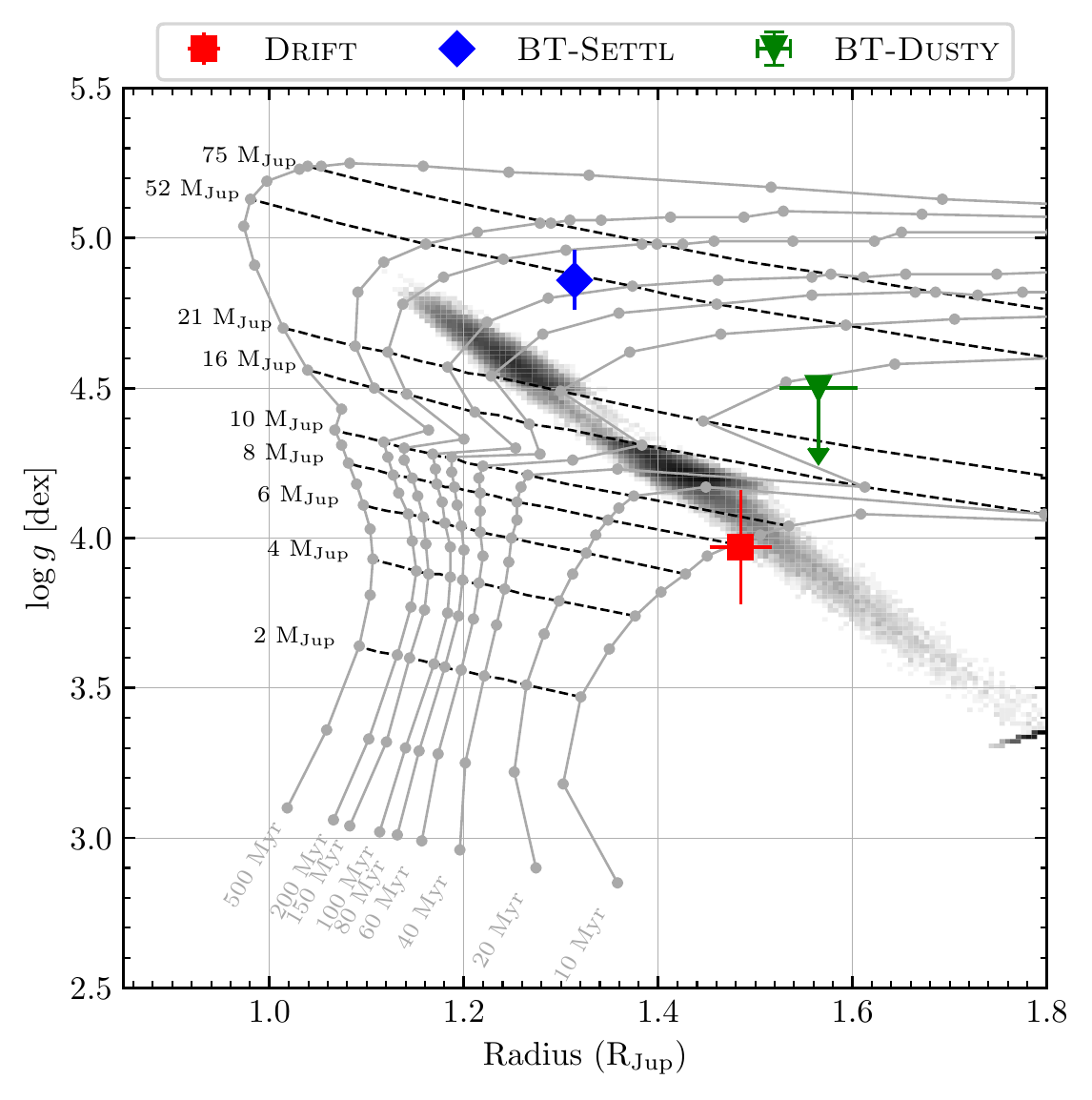}
    \caption{Radius-gravity diagram comparing the best fit atmospheric models to isochrones (gray solid lines) and mass tracks (black dashed lines) from the COND03 evolutionary model \citep{Baraffe2003}. The radius and gravity derived via a Monte Carlo error propagation from the luminosity and age given in \citep{Currie2018} is also shown for comparison (grayscale two-dimensional histogram), plotted on a logarithmic color scale to highlight the isoluminosity contour consistent with the measured luminosity of $\kappa$~And~b \citep{Currie2018}.}    \label{fig: Kap And COND03}
\end{figure}

We repeated this exercise on an interpolated version of each grid to search for a better fit with combinations of $T_{\rm eff}$ and $\log g$ not included within the original grid. We constructed a new grid of models with an arbitrarily small grid spacing of $\Delta T_{\rm eff}=1$\,K and $\Delta \log g=0.01$\,[dex]. Models were constructed by performing a bilinear interpolation of the logarithm of the flux calculated within the seven photometric bands listed in Table~\ref{Photometry summary Kap And} and the sixteen spectral channels of the CHARIS spectrum presented in \citet{Currie2018}. We find a significantly reduced $\chi^2_{\nu}$ of 1.2 (compared with 1.7 in the coarse grid) for the {\sc Drift-Phoenix} model at $T_{\rm eff}=1739$\,K and $\log g=4.0$\,[dex] (Figure~\ref{fig:model_plot}, bottom panel). A similar reduction in $\chi^2$ is seen for the other two grids. The $\chi^2$ surface for the interpolated version of the three best-fitting grids is shown in Figure~\ref{fig:model_chi2}, showing the radius required to minimize $\chi^2$ given the distance of 50.0\,pc (Table \ref{Kappa And parameters}), and the credible regions derived from the $\Delta\chi^2$ (with no treatment for model uncertainties).

We find that the best-fit models are able to reproduce the observed SED, and are consistent with one another, over the $JHK$ range. At shorter and longer wavelengths where the uncertainties on the photometric measurements are larger the models diverge slightly. The lower-gravity {\sc Drift-Phoenix} model significantly under-predicts the flux at $Y$, while slightly over predicting the flux at $M^{\prime}$. The higher-gravity {\sc BT-Settl} model predicts a larger flux at $Y$, consistent with measured flux, but significantly under-predicts the flux at $L^{\prime}$. Due to the differences in treatment for cloud formation and sedimentation within these models, as well as revisions to opacity tables used to compute the emergent spectra, it is difficult to ascribe the differences between the best fit models to a particular property or feature of the models. A future study that incorporates high resolution spectroscopy and precision photometry between 1--5\micron\ in conjunction with a retrieval-based modeling approach will allow us investigate the effect of the bulk (e.g., temperature, surface gravity, luminosity) and photospheric (e.g., cloud extent and vertical distribution, dust condensation and sedimentation) properties on the emergent spectra of this object.

Figure \ref{fig: Kap And COND03} shows how the best-fit radii and gravities derived from atmospheric modeling compare to predictions from luminosity evolution models for a given age and mass.   The grey contours adopt $\kappa$ And b's luminosity derived from \citet{Currie2018} ($\log L/L_{\odot} = -3.81 \pm 0.05$) and an age range of $47\pm30$\,Myr -- similar to the age range derived from a CHARA radius measurement of the host star in \citet{Jones2016}.   As the best fit gravity for the BT-DUSTY model is at the lower limit of the grid (log(g) $\sim$ 4.5) we display its point with a downward arrow; {\sc Drift-Phoenix} and {\sc BT-Dusty} model parameters are shown with error bars corresponding to the 68\% confidence interval.

The best-fitting model atmosphere fit -- {\sc Drift-Phoenix} -- implies a radius and gravity consistent with evolutionary model predictions for an age of $t$ $\le$ 40 Myr, yielding a mass nominally of 10 $M_{\rm Jup}$ and less than 20 $M_{\rm Jup}$ considering errors.  The {\sc BT-Dusty} model implies a mass less than $\sim$ 30-35 $M_{\rm Jup}$; its radius/gravity is inconsistent with evolutionary models but could be reconciled if the gravity is lower by 0.5 dex or radius smaller by 0.2 R$_{\rm Jup}$, either of which would imply a mass less than 20 $M_{\rm Jup}$.   The best-fit {\sc BT-Settl} model's radius and gravity imply higher masses and far older ages which are consistent with the early analysis by \citet{Hinkley2013}.  However, the implied radii and gravities are inconsistent with predictions from evolutionary tracks in Figure \ref{fig: Kap And COND03}.   They also imply ages significantly older than and thus inconsistent with ages derived from $\kappa$ And A's radius using CHARA interferometry \citep{Jones2016}.

\section{Orbital Fitting} \label{sec: Astrometry}
Astrometric monitoring of $\kappa$ And b over eight years helps constrain orbital motion of the $\kappa$ And system.
Relative positions of $\kappa$ And b obtained by Subaru/Keck observations are summarized in Table \ref{Astrometry Kap And b}.
\cite{Blunt2017} estimated orbital parameters of $\kappa$ And b from only three relative positions from 2011--2012 \citep{Carson2013}, which correspond to the change in a position angle (PA) of $\Delta$PA$\sim$0.4$^\circ$.
\cite{Currie2018} observed relative positions of $\kappa$ And b in 2017 and derived orbital parameters of $\kappa$ And b from astrometric data prior to 2013 and their results ($\Delta$PA$\sim$5.5$^\circ$).
We re-analyzed orbital motion of $\kappa$ And b, using relative positions of $\kappa$ And b obtained by Subaru/HiCIAO+SCExAO in 2016 and Keck/NIRC2 in 2018. The position angle change between the first Subaru/HiCIAO report and the latest NIRC2 data is $\sim$7$^\circ$.

ExoSOFT \citep[][]{Mede2017} was used for orbital fitting, which takes advantage of a several techniques, including the Markov Chain Monte Carlo (MCMC) approach, to estimate dynamical parameters from relative positions at different epochs.
First, we used two modules that are incorporated in ExoSOFT: simulated annealing to search for the global minimum and sigma tuning (ST) to determine reasonable step sizes.
Finally, we ran {\it emcee} mode (an MCMC ensemble sampler \citep{emcee}) with $n=6\times10^8$ total samples across 500 walkers to fit the orbit of $\kappa$ And b and to estimate its dynamical and orbital parameters.  We adopted 19.3--20.7 mas for a parallax range and 2.65--2.95 $M_{\odot}$ for a mass range of the $\kappa$ And system during the final fitting, as ExoSOFT does not currently offer those parameters to remain fixed when running in the {\it emcee} mode. 
The samples for the parameters $(e, P, T_0, i,\Omega, \omega$) were drawn from uniform proposal distributions. The priors for $e, T_0, \Omega$ and $\omega$ were set as uniform, while we assumed a Jeffrey's prior function for the semi-major axis ($a^{-1}/\ln{a_{\rm max}/a_{\rm min}}$), with $i$ and $P$ given the priors $p(i) \propto \cos(i)$ and $p(P) \propto 1/P$, respectively.  
Providing only direct imaging data, orbital fitting using ExoSOFT finds the total mass of the $\kappa$ And system ($m_{\rm total}$), although it is capable of solving for the individual masses when coupled with radial velocity data \citep[see Section 2 of][for more details]{Mede2017}.

\begin{figure*}
    \begin{tabular}{cc}
    \begin{minipage}{0.5\hsize}
    \centering
    \includegraphics[scale=0.32]{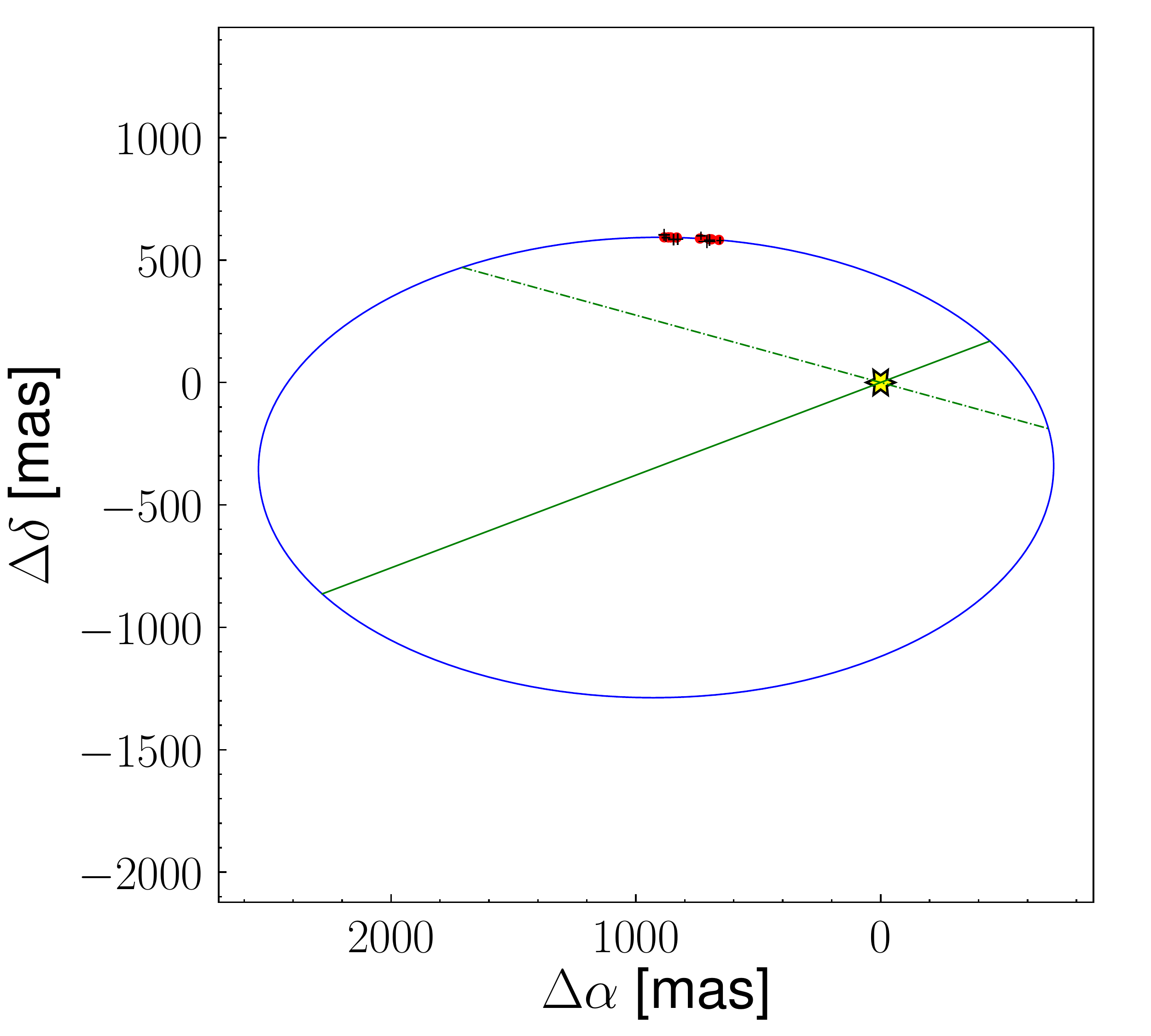}
    \end{minipage}
    \begin{minipage}{0.5\hsize}
    \includegraphics[scale=0.32]{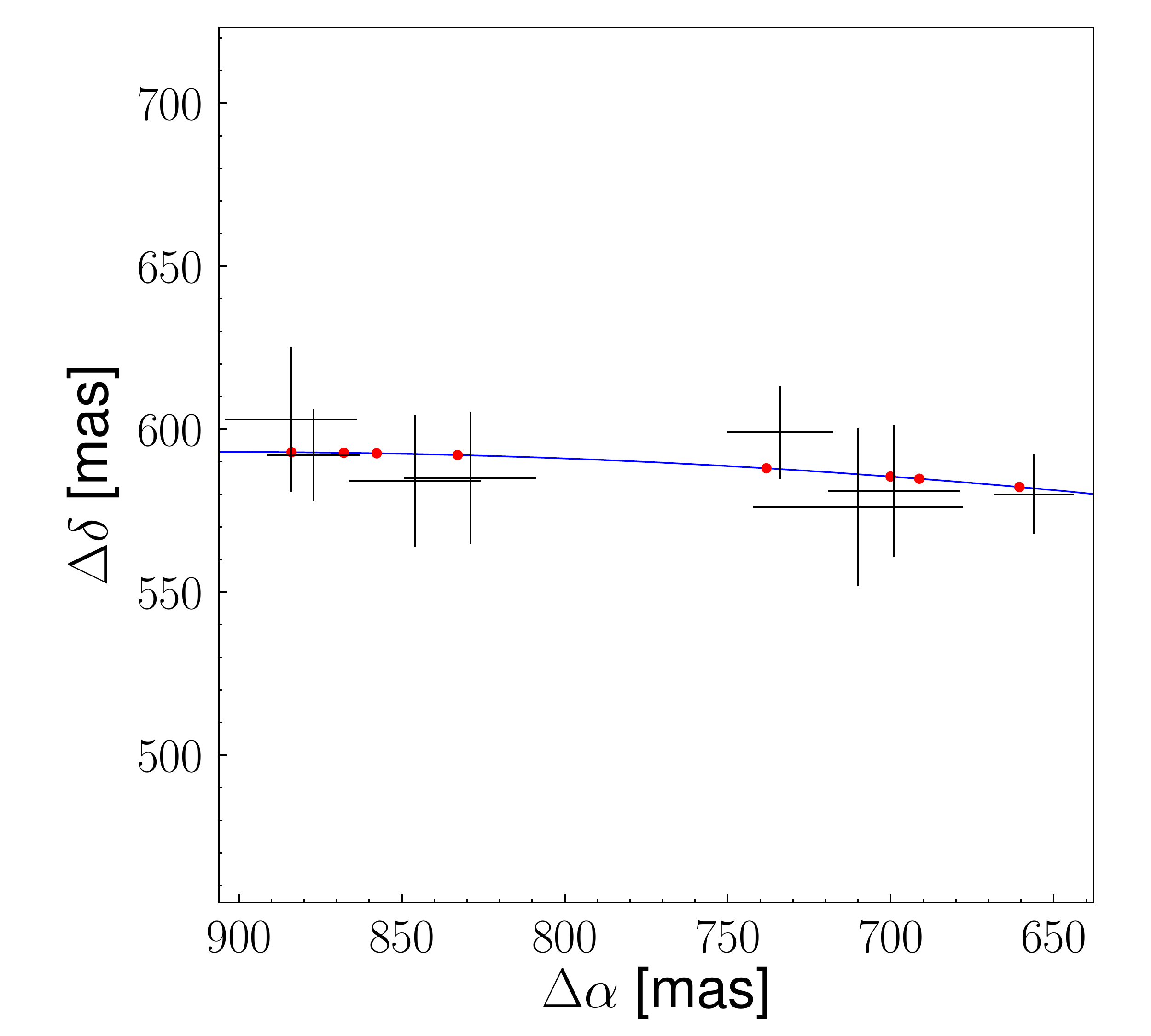}
    \end{minipage}
    \end{tabular}
    \caption{Orbital fitting of $\kappa$ And b with ExoSOFT. A blue ellipse is the best-fit solution for the orbit of $\kappa$ And b, where $\kappa$ And b moves clockwise:(left) the full orbit and (right) a zoom-in view near the current positions. Black crosses are relative positions of $\kappa$ And b obtained by previous Subaru/Keck observations, red plots are predicted locations of the best-fit orbit at each epoch. The solid and dashed lines in the left figure correspond to the projected semi-major axis and the line of nodes, respectively.}
    \label{ExoSOFT fitting Kap And}
\end{figure*}

\begin{figure}
\centering
\includegraphics[width=0.75\columnwidth]{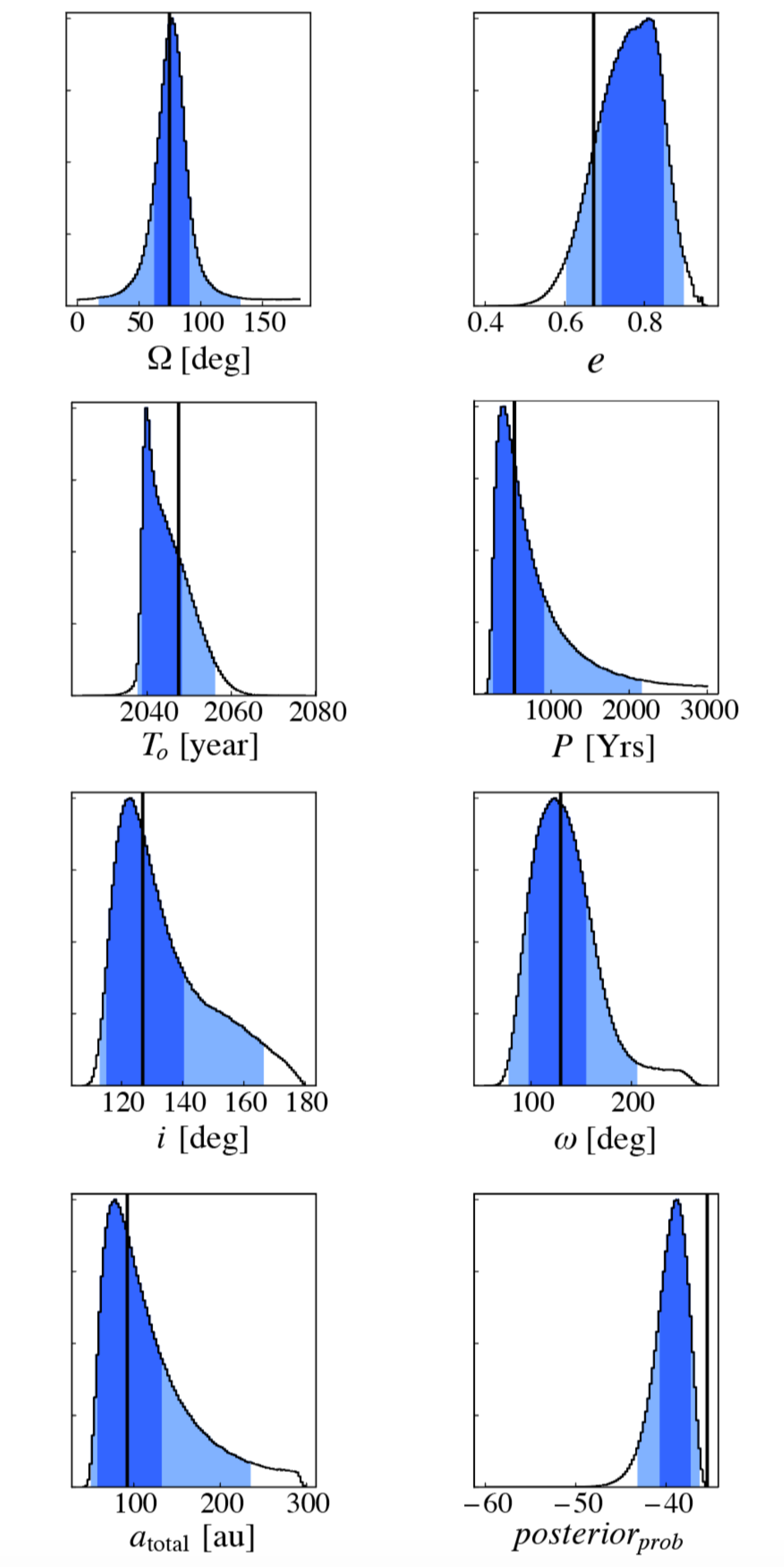}
\caption{Posteriors of MCMC parameters used in ExoSOFT for the $\kappa$ And system. Dark and light blue {\bf regions} correspond to 1 and 2 $\sigma$, respectively. Solid black lines represent best-fit values of each parameter (where 'best-fit' refers to the orbital parameter set with the lowest $\chi^2$ value).}
\label{ExoSOFT summary Kap And}
\end{figure}

Figure \ref{ExoSOFT fitting Kap And} shows a result of the orbital fitting with ExoSOFT. Posteriors of the parameters used in ExoSOFT are shown in Figure \ref{ExoSOFT summary Kap And}.
The mass ratio between the companion and the central star is $q\sim$0.005, namely $m_{\rm total}\sim m_{\rm star}$.
If the posterior function of $m_{\rm total}$ follows a Gaussian, we can estimate the dynamical mass of $\kappa$ And A, which is independent of previous photometric/spectroscopic studies.
However, our calculation could not robustly constrain $m_{\rm total}$ due to the limited number of $\kappa$ And b locations.
Our results of other orbital parameters achieved a best fit with a reduced $\chi^2$ of 0.958 and are in good agreement with the previous report in \cite{Currie2018}.  In the ExoSOFT fit the least convergent parameter was that of $P$ having an integrated autocorrelation time of 921, equating to $6.5\times10^5$ effective samples.
Astrometric monitoring for the next ten years is required to more accurately determine the orbital parameters of the $\kappa$ And system.

\begin{table*}
\caption{Orbital parameters of $\kappa$ And b}
\centering
\begin{tabular}{cccc} \\ \hline\hline
Parameter & Median & 68\% confidence level & 95\% confidence level \\ \hline
$a_{\rm tot}$ [au]  & 103.6 & [57.4, 133.4] &[50.3, 236.0] \\
$P$ [yr] & 631.1 & [242.4, 900.4] & [198.6, 2148.9] \\
$e$ & 0.77 & [0.69, 0.85] & [0.60, 0.90] \\
$i$ [deg] & 130.0 & [114.9, 140.0] & [112.6, 166.6] \\
$\omega$ [deg] & 130.7 & [96.6, 155.4] & [77.0, 205.0] \\
$\Omega$ [deg] & 76.5 & [61.3, 90.5] & [16.4, 132.1] \\
$T_0$ [yr] & 2044.1 & [2038.4, 2047.9]  &[2037.5, 2056.3] \\ \hline
\end{tabular}
\label{orbital fitting results Kap And}
\end{table*}

\section{Discussion} \label{sec: Discussions}
    \subsection{Formation and Evolution Scenario} \label{sec: Formation}

Our atmospheric modeling favors 1700--1900K, a surface gravity of log(g)$\sim$4.0-4.5, and a radius of 1.3–-1.6 $R_{\rm Jup}$ with cloudy atmosphere.
The best-fit model (the {\sc drift-phoenix} model) is consistent with $\leq40$ Myr and $<20 M_{\rm Jup}$ in the evolutionary model.
$\kappa$ And b is a good laboratory for understanding formation and an early stage of evolution of gas giant/low-mass brown dwarf.

We reconfirmed that $\kappa$ And b is likely to have a larger eccentricity and semi-major axis than GJ 504 b \citep{Bonnefoy2018} and HR 8799 b,c,d,e \citep[][]{Wang2018}.
$\kappa$ And b may have experienced a strong excitation of the eccentricity by gravitational interactions between neighboring planets such as planet-planet scattering.
Planetesimal accretion and accumulation of a disk gas cannot pump up the eccentricity of a planet's orbit up to $\sim 0.8$.
In fact, a wide orbit of $\kappa$ And b cannot be reconciled with in-situ core accretion scenario. Although the minimum core mass for gas giant formation requires only a few Earth masses at $\sim$100 au \citep{Piso2014}, the core growth at 100 au takes a much longer time than the estimated age of the $\kappa$ And system.
\cite{Bonnefoy:2014dx} proposed another possible formation scenario for $\kappa$ And b (i.e., a hot-start model); it may have formed via gravitational instability at almost the same orbital separation as the current location.

It may also be possible that $\kappa$ And b was scattered to its current location \citep[e.g.][]{2002Icar..156..570M,2008ApJ...686..621F,Nagasawa2008}.
Since the age of $\kappa$ And A was estimated to be $\sim$40--50\,Myr, dynamical instability was likely to have occurred if three or more giant planets co-existed in an outer region. An outwardly-scattered planet, namely $\kappa$ And b, can remain on a highly eccentric orbit because of less efficient/no dynamical frictions damping of the eccentricity \citep{Muto2011}.
To investigate this scenario, we consider that a planet-planet scattering event occurred after disk dispersal.
The planet-planet scattering requires close encounters of planets, which are induced easily in a system of three or more planets. The behaviors of planet-planet scatterings which are involved in more than three planets need to be numerically examined by N-body simulations. In this study, we discuss a simple case with three giant planets.
We assume i) three massive gas giants/brown dwarfs on nearly coplanar, circular, and tightly packed orbits around $\kappa$ And, ii) one of them is ejected from the system, iii) $\kappa$ And b is the outer planet of two remaining objects, iv) the ejected planet has a smaller mass than $\kappa$ And b \citep[as shown by N-body simulations of planet-planet scatterings;][]{2002Icar..156..570M}, and v) the three objects have similar radii.
Under these assumptions, we infer the mass and orbital elements of an unseen (potential) planet in the $\kappa$ And system.

After dynamical instability happened, the eccentricity of an outer remaining object ($\kappa$ And b) is determined by
\begin{eqnarray}
e_{\rm out} \simeq \frac{m_{\rm in}}{m_{\rm out}}\times\sqrt{\frac{m_{\rm out}+m_{\rm eje}}{m_{\rm out}+m_{\rm in}}},
\label{eq: ecc}
\end{eqnarray}
where $m$ corresponds to the mass of an object and the subscripts of "in", "out", and "eje" correspond to the inner, outer ($\kappa$ And b), and ejected objects, respectively \citep[][]{Ida2013}.
Using Equation (\ref{eq: ecc}) and the mass and eccentricity of $\kappa$ And b, i.e., $m_{\rm out}=13 M_{\rm Jup}$ and $e_{\rm out}=0.77\pm0.08$,
we can estimate the mass of the inner object as a function of the mass of the ejected object (see Table \ref{mass scattering}). We note that the error bar shown in Table \ref{mass scattering} comes from only the error of eccentricity ExoSOFT provided. Estimating $\kappa$ And b's mass depends on the age and the evolutionary models largely and we do not include this error.
With these assumptions, the potential inner companion (planet) has mass of $m_{\rm in}\gtrsim10 M_{\rm Jup}$.
We note that Equation (\ref{eq: ecc}) is not applicable to the case where $\kappa$ And had initially four or more giant planets in an outer region because orbital evolution of such a system cannot be described analytically any longer.

\begin{table}
\caption{Mass estimation of a potential inner companion around $\kappa$ And}
\centering
\begin{tabular}{ccc} \\ \hline\hline
ejected object [$M_{\rm Jup}$] & inner object [$M_{\rm Jup}$] \\ \hline
2 & $13.2^{+1.9}_{-1.7}$ \\
4 & $12.2^{+1.7}_{-1.6}$ \\
6 & $11.3^{+1.6}_{-1.5}$ \\
8 & $10.6\pm1.4$ \\
10 & $10.0^{+1.4}_{-1.3}$ \\ \hline
\end{tabular}
\label{mass scattering}
\end{table}

Since no point source other than $\kappa$ And b is seen in Figure \ref{SCExAO 2016}, 
we discuss the mass limit of a detectable planet around $\kappa$ And.
The latest SCExAO+CHARIS observation reached a better contrast limit in the wavelength-collapsed image \cite[][]{Currie2018}: $\sim$15$M_{\rm Jup}$, $\sim$8--10$M_{\rm Jup}$, and $\sim$3--5$M_{\rm Jup}$ at 12.5, 25, and 50\,au, respectively, using a hot-start model \citep[COND03;][]{Baraffe2003}.
With the deepest contrast limits around $\kappa$ And, SCExAO+CHARIS observations can suggest that an inner companion can be located at $\lesssim$ 25\,au.

Combining radial velocity (RV) methods with direct imaging enables us to give stringent constraints on orbital parameters of a substellar-mass companion \citep[e.g.,][]{Calissendorff2018,Bonnefoy2018}. The lack of absorption lines obscures precise RV measurements of massive stars such as $\kappa$ And A ($\sim$B9 star) due to high temperature and rapid rotation. In fact, archival RV observations reported large errors $>$1 km/s \citep[][]{Hinkley2013,Becker2015}. 
Host-star astrometry is also useful, but estimating accurate acceleration of such a bright star by a combination of {\it Gaia} and {\it Hipparcos} telescopes cannot avoid systematic errors between these telescopes \citep[][]{Brandt2018catalogue}.
Accumulating {\it Gaia} data sets will possibly help to measure the dynamical mass of $\kappa$ And b in the future.

\subsection{Future Work} \label{sec: Future Work}
Spectral features of substellar-mass objects within $\sim$1--5 $\mu$m depend on molecular absorption such as FeH, H$_2$O, K I, CH$_4$, and CO. Effective temperature, surface gravity, or C/O ratio parameters affect IR spectrum \citep[e.g.,][]{Sorahana2012,Sorahana2014}. Our study uses only photometry and low-resolution spectroscopy, which can induce degeneracy between $T_{\rm eff}$ and $\log g$ and the best-fit objects for the field-gravity objects in Figure \ref{fig:bestfit_library}.
Although a precise determination of the gravity of $\kappa$~And~b will require higher spectral resolution observations, our measurements demonstrate that the object likely has a low surface gravity when considering the age of the system, consistent with the planetary mass predicted from a comparison with evolutionary models (e.g., \citealp{Currie2018}).
For future work, as introduced in \cite{Currie2018}, higher-resolution spectroscopy helps to investigate $\kappa$ And b's atmosphere in detail.
Subaru/CHARIS has another spectroscopic mode with high-resolution (R$\sim$65-75) in $J$, $H$, and $K$ bands\footnote{\url{https://scholar.princeton.edu/charis/capabilities}}. 
Keck/OSIRIS could extract HR8799 b's spectrum with higher resolution \citep[R=4000;][]{Barman2015,Petit2018}. 
A mid-spectral-resolution integral field unit (IFU) combined with AO has the capability to extract the detailed spectrum and to investigate atmospheric/evolutionary mechanisms of $\kappa$ And b as mentioned in Section \ref{sec: Comparison with Spectral Templates}.
Furthermore, mid-IR (MIR) wavelength photometry/spectroscopy will also provide useful information.
JWST/MIRI is expected to obtain untouched atmospheric parameters of exoplanets at MIR such as NH$_3$, CH$_4$, H$_2$O, CO$_2$, and PH$_3$ \citep[][]{Danielski2018}. Combining these follow-up observations will provide improved models for $\kappa$ And b.

We also investigate the possibility to detect a potential inner planet.
Radial velocity and host-star astrometry are more sensitive to close-in planet than direct imaging.
However, as mentioned in Section \ref{sec: Formation}, it is difficult for these methods to search for inner planets around $\kappa$ And.
As we could not constrain an inclination of the potential inner planet, transit observation is almost a blind search.
Future high-contrast imaging instrument with a better contrast level and inner working angle, e.g., Thirty Meter Telescope (TMT), will help to search for inner planets and to promote orbital evolution mechanisms of $\kappa$ And b. Continuing direct imaging with current ground-based telescopes also helps to add further plots of $\kappa$ And b for better orbital fitting.

\section{Conclusion} \label{sec: Conclusion 2}

We used Subaru/SCExAO+HiCIAO and Keck/NIRC2 to investigate $\kappa$ And b's SED and to fit the orbit by gathering our results and previous high-contrast iamging studies.
We detected $\kappa$ And b with SNRs of $\sim$130 and 10 in the HiCIAO $H$- and $Y$-band, and $\sim$13 in the NIRC2 $K_{\rm s}$-band, respectively. 
The $Y$-band photometry was combined with previous photometric/spectroscopic studies for an empirical comparison with spectral templates and for synthetic SED modeling with atmospheric models.
Empirical comparisons showed that $\kappa$ And b is likely a low-gravity object, albeit one with a slightly wide range of plausible spectral types than previously inferred (L0-L2 instead of L0-L1).
We also investigate gravitational scores of the library objects and found that the best-fit objects may give lower gravity than previously reported\footnote{Some intermediate gravity dwarfs also provide good fits to $\kappa $ And b spectrum.  However, we did not take into account information about the system’s age in our fitting (i.e. we did not impose a ‘prior’ on the gravity classification of ‘b’ given the age of the primary).   Doing so would have even more strongly favored low gravity objects.}.
The best fit among used models is the {\sc Drift-Phoenix} model at $T_{\rm eff}$=1700 K and $\log g$=4.0 [dex]. With the interpolated grid the best fit is located at $T_{\rm eff}$=1739 K and $\log g$=4.0 [dex].
More than 7 years have passed since the first report of $\kappa$ And b in 2011 January, which resulted in a position angle change of ${\rm PA}\sim$ $7^\circ$.
By running ExoSOFT we found that the orbit is likely highly eccentric, which suggests a possibility that $\kappa$ And b has experienced orbital migration due to planet-planet scattering.
Our detection limit could partially set a constraint on the existence of a potential inner companion.
Previous studies except \citet{Currie2018} discussed the formation and evolution mechanisms of $\kappa$ And b without taking into account of its eccentricity. Our analysis will be help to update the synthetic understanding of formation and evolution mechanism of the $\kappa$ And system.

For future work, spectroscopic studies with higher resolution such as high-resolution mode of Subaru/CHARIS or Keck/OSIRIS help to investigate $\kappa$ And b's atmosphere in details.
JWST will enable to obtain spectral/photometric information at MIR.
TMT is expected to achieve higher contrast enough to detect inner planetary-mass object and update the orbital discussions.
Our work motivates follow-up observations for future telescopes and further discussions of formation/evolution mechanisms of $\kappa$ And b.

\acknowledgments

The authors would like to thank the anonymous referees for their constructive comments and suggestions to improve the quality of the paper.
This paper is based in part on data collected at Subaru telescope and obtained from the SMOKA, which is operated by the Astronomy Data Center, National Astronomical Observatory of Japan.
Some of the data presented herein were obtained at the W. M. Keck Observatory, which is operated as a scientific partnership among the California Institute of Technology, the University of California and the National Aeronautics and Space Administration. The Observatory was made possible by the generous financial support of the W. M. Keck Foundation.
This work presents results from the European Space Agency (ESA) space mission Gaia. Gaia data are being processed by the Gaia Data Processing and Analysis Consortium (DPAC). Funding for the DPAC is provided by national institutions, in particular the institutions participating in the Gaia MultiLateral Agreement (MLA). The Gaia mission website is \url{https://www.cosmos.esa.int/gaia}. The Gaia archive website is \url{https://archives.esac.esa.int/gaia}.

TU acknowledges JSPS overseas research fellowship. 
This work was supported by JSPS KAKENHI Grant Numbers JP17J00934, 15H02063, and 18H05442. 
T.C. was supported by a NASA Senior Postdoctoral Fellowship and NASA/Keck grant LK-2663-948181; R.D.R. was supported by NASA grant NSSC17K0535.
The development of SCExAO was supported by JSPS (Grant-in-Aid for Research \#23340051, \#26220704 \& \#23103002), Astrobiology Center of NINS, Japan, the Mt Cuba Foundation, and the director's contingency fund at Subaru Telescope.
CHARIS was developed under the support by the Grant-in-Aid for Scientific Research on Innovative Areas \#2302.

The authors wish to acknowledge the very significant cultural role and reverence that the summit of Mauna Kea has always had within the indigenous Hawaiian community.  We are most fortunate to have the opportunity to conduct observations from this mountain.

\appendix
\section{HiCIAO Filter Transmission} \label{sec: HiCIAO Filter Transmission}
\begin{figure}[h]
\centering
\includegraphics[scale=0.35]{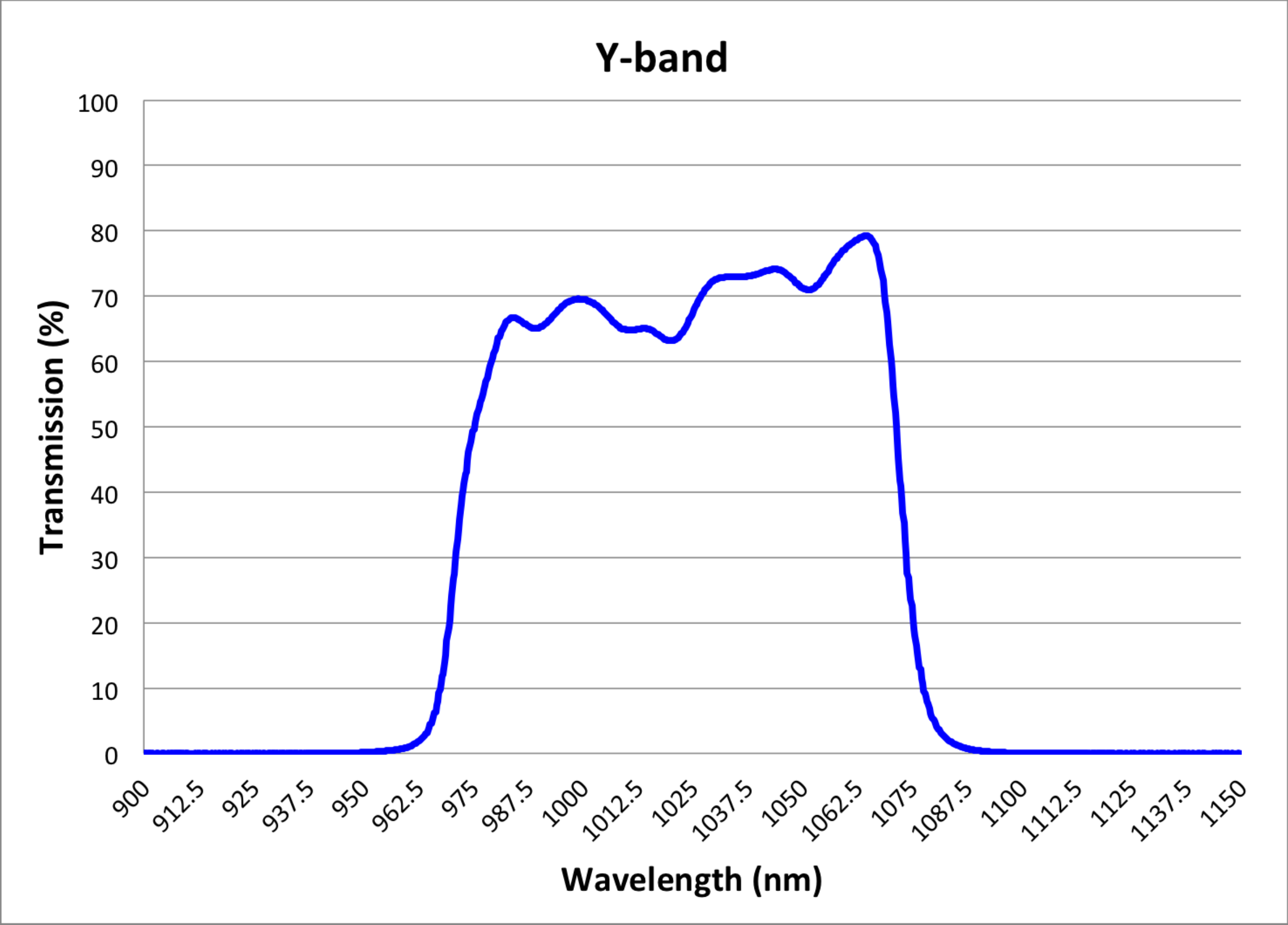}
\caption[HiCIAO $Y$-band transmission]{Transmission of the $Y$-band test filter under 77K.}
\label{Y-band transmission}
\end{figure}

In Figure \ref{Y-band transmission} we show the transmission of a $Y$-band test filter that has almost the same specifications as HiCIAO.
HiCIAO observations were basically carried with optical bench temperature of $\sim$80 K and this filter transmission is measured under 77 K.

\bibliographystyle{aasjournal}                                                              
\bibliography{library}                                                                

\end{document}